\documentclass[english, 12pt]{article}
\usepackage[T1]{fontenc}
\usepackage[utf8]{inputenc}
\usepackage[english]{babel}
\usepackage{lmodern}
\usepackage[top=1in, bottom=1in, left=1in, right=1in]{geometry}
\usepackage{setspace}
\usepackage[pdftex]{graphicx}
\usepackage{verbatim}
\usepackage[width=0.8\textwidth]{caption}
\usepackage{amsmath,amssymb,amsopn,dsfont,mathrsfs,bm}
\usepackage{float,placeins,flafter,longtable,array,booktabs}
\usepackage{color}
\usepackage{natbib}
\usepackage{xr}
\usepackage[hidelinks]{hyperref}
\makeatletter

\newtheorem{thm}{Theorem}

\newtheorem{prop}{Proposition}

\newtheorem{hyp}{Assumption}

\newcommand{\st}[1]{\texttt{#1}}
\newcommand{\ind}[1]{1\left\{#1\right\}}
\newcommand{\AVSQ}{\text{AVSQ}}
\newcommand{\ACE}{\text{ACE}}

\newcommand{\DID}{\text{DID}}
\newcommand{\0}{\bm{0}}
\newcommand{\1}{\bm{1}}
\newcommand{\Dnc}{\mathcal{D}^{\text{nc}}}
\newcommand{\Dl}{\mathcal{D}^\ell}
\newcommand{\Dr}{\mathcal{D}^{\text{r}}_1}
\newcommand{\R}{\mathbb R}

\newcommand{\eps}{\varepsilon}

\newcommand{\Cov}{\text{Cov}}
\newcommand{\Supp}{\text{Supp}}

\newcommand{\sgn}{\text{sgn}}

\allowdisplaybreaks

\date{}

\begin{document}
	
	\title{Treatment-Effect Estimation in Complex Designs under a Parallel-trends Assumption\thanks{We would like to thank Henri Fabre for his excellent research assistance, St\'ephane Bonhomme and Jonathan Roth for their comments and Ryo Okui for his discussion.}}
	\author{Cl\'ement de Chaisemartin\thanks{Sciences Po Paris, clement.dechaisemartin@sciencespo.fr} \and Xavier D'Haultf\oe uille\thanks{CREST-ENSAE, xavier.dhaultfoeuille@ensae.fr.}}
	
	\maketitle
	
	\begin{abstract}
This paper considers the identification of dynamic treatment effects with panel data, in complex designs where the treatment may not be binary and may not be absorbing. We first show that under no-anticipation and parallel-trends assumptions, we can identify event-study effects comparing outcomes under the actual treatment path and under the status-quo path where all units would have kept their period-one treatment throughout the panel. Those effects can be helpful to evaluate ex-post the policies that effectively took place, and once properly normalized they estimate weighted averages of marginal effects of the current and lagged treatments on the outcome. Yet, they may still be hard to interpret, and they cannot be used to evaluate the effects of other policies than the ones that were conducted. To make progress, we impose another restriction, namely a random coefficients distributed-lag linear model, where effects remain constant over time. Under this model, the usual distributed-lag two-way-fixed-effects regression may be misleading. Instead, we show that this random coefficients model can be estimated simply. We illustrate our findings by revisiting \cite*{gentzkow2011}.

\textbf{Keywords:} difference-in-differences, complex design, random coefficients models.
\end{abstract}

\section{Introduction} 
\label{sec:introduction}

Most work on difference-in-differences has focused on simple designs, namely classical designs where some units receive a binary treatment at the same time whereas other do not, or staggered adoption designs, where units begin to receive an absorbing binary treatment at different points in time, or heterogeneous adoption designs where all units are initially untreated before the treated units start receiving heterogeneous doses. However, more complex designs are actually prevalent. In previous work \citep{de2020difference}, we conducted a survey of the 100 papers with the most Google Scholar citations published by the American Economic Review from 2015 to 2019. Of those, 26 use a two-way fixed effects (TWFE) regression to estimate the effect of a treatment on an outcome.\footnote{Counting TWFE regressions is in line with the long-held belief that this regression is the uncontroversial treatment-effect estimator in a DID research design.} Of these 26 papers, only two have a classical design, four have an absorbing and binary treatment with variation in treatment timing, and two have an heterogeneous adoption design. The remaining 18 have a more complex design, with a non-binary and/or non-absorbing treatment, and where units may receive heterogeneous treatment doses even at period one.

\medskip
The aim of this paper is to study what can be learnt in such designs. We first focus on identification under no-anticipation and parallel-trends assumptions, summarizing and extending results from \cite{de2020difference}. A first challenge is that the usual parallel-trends condition, where potential-outcome trends without treatment are mean independent of the treatment path, may not have much identification power if many units actually receive a non-zero dose of treatment at period one. We consider another parallel-trends condition, which accommodates this issue. It requires that conditional on their period-one treatment, outcome trends in the status-quo counterfactual where units keep their period-one treatment are mean independent of units' treatment path. Pre-trends tests can be used to assess the plausibility of this parallel-trends assumption. Conditioning on units' period-one treatment is important: without that conditioning, our parallel-trends assumption would actually rule out effects of the lagged treatments on the outcome, once combined with the usual parallel-trends assumption.

\medskip
Under our parallel-trends condition, to identify the effect of having departed from one's status-quo treatment $\ell$ periods ago, we can form a difference-in-difference estimand comparing units that have first switched treatment $\ell$ periods ago with units that have not switched yet and that had the same treatment at period one. The corresponding event-study effects may be more challenging to interpret than in staggered adoption designs. Nonetheless, we show that they can be useful to do an ex-post cost-benefit analysis of the policies that effectively took place over the study period. Moreover, once properly normalized they estimate weighted averages of marginal effects of the current and lagged treatments on the outcome. Finally, they can be used to test relevant null hypotheses, such as whether lagged treatments affect the outcome or not. The corresponding estimators are computed by the \st{did\_multiplegt\_dyn} Stata, R, and Python commands.

\medskip
Still, these event-study effects cannot be used to separate the marginal effects of the current treatment and of some specific treatment lags. Moreover, they cannot be used to evaluate alternative policies, or determine the optimal policy. To make progress on these important questions, we consider in the second part of the paper a further restriction, in the form of a random coefficients distributed-lag linear model. This model is restrictive, but allows for heterogeneity in treatment effects that can be correlated with the treatment path. We first decompose the usual distributed-lag TWFE regressions under this model, and show that they usually fail to identify convex combinations of unit-specific effects. Then, we show that expectations of the random coefficients can be identified and estimated simply and without tuning parameters if the treatment variable takes a finite number of values. The corresponding estimators are computed by the \texttt{dist\_lag\_het} R package, available at \url{https://github.com/chaisemartinPackages/dist_lag_het}.

\medskip
Finally, we illustrate our theoretical results by studying the effect of newspapers on electoral turnout in the US between 1868 and 1928, revisiting \cite*{gentzkow2011}. We extend their analysis by considering a dynamic rather than static setup, thus allowing the number of past newspapers to potentially affect current turnout. Our event-study effects, relying only on no-anticipation and parallel trends assumptions, show a large positive effect of newspapers on turnout. Though these parameters cannot bring a definitive answer to this question, they also suggest that the current number of newspapers has a larger effect than the lagged number of newspapers, and our estimates of the random coefficients distributed lag model confirm this finding. On the other hand, the distributed-lag TWFE regression leads to the opposite conclusion: the coefficient on lagged newspapers is positive and significant, while the coefficient on current newspapers is small and insignificant. Our decomposition of those coefficients shows that they may be unreliable if effects vary across counties.

\medskip
This paper contributes to a rapidly growing literature on difference-in-differences (DID). In that literature, most papers have focused on designs with a binary and absorbing treatment \citep{dcDH2020,callaway2018,abraham2018,borusyak2020revisiting} or on heterogeneous adoption designs where all units are untreated at period one before treated units start receiving heterogeneous doses \citep{fricke2017identification,dcDH2020,callaway2021difference,de2022heterogeneousadoption}. The first part of the paper summarizes and extends results from \cite{de2020difference}, that can accommodate complex designs with a non-binary and/or non-absorbing treatment, and where units may receive heterogeneous treatment doses even at period one. We emphasize that under a parallel-trends assumption alone, we can identify weighted averages of the effect of the current and lagged treatments on the outcome, without being able to separately estimate those effects. This motivates our analysis of the random coefficients distributed-lag model in the second part of the paper. There, we show that the work of \cite{chamberlain1992efficiency}, \cite{arellano2012identifying}, and \cite{graham2012identification} can fruitfully be applied to estimate distributed-lag models with heterogeneous effects. Compared to these papers, we highlight the fact that averages of the random coefficients can be estimated simply and without tuning parameters if the treatment variable takes a finite number of values. We also provide identification conditions for the average of each random coefficient, rather than the average of the whole vector. 

\medskip
The paper is organized as follows. Section 2 presents the setup, our main assumptions and the type of designs we consider. Section 3 considers identification and estimation under no-anticipation and parallel-trends assumptions. Section 4 considers identification and estimation when one also imposes a random coefficients distributed-lag linear model. Section 5 applies the results of Sections 3 and 4 to estimate the effect of newspapers on electoral turnout. Proofs are relegated to the appendix.


\section{Setup} 
\label{sec:setup_assumptions_and_parameters_of_interest}

\subsection{Treatment and potential outcomes} 
\label{sub:setup_and_assumptions}

We consider a panel of $G$ groups observed at $T$ periods, respectively indexed by $g$ and $t$. Groups can be locations, like states, counties, or municipalities, but could also just be individuals or firms. We seek to identify (dynamic) effects of a treatment on a given outcome. Let $D_{g,t}$ denote the treatment of group $g$ at period $t$, with support denoted by $\mathcal{D}_t$, supposed to be independent of $g$. Also, let $\bm{D}_g=(D_{g,1},...,D_{g,T})$ denote the treatment path of group $g$, and let $\mathcal{D}$ denote its support, also assumed to be independent of $g$. We let $F_g$ be the first date at which $g$ switches treatment: $F_g= \min \{t\geq 2: D_{g,t}\ne D_{g,1}\}$. If $g$ never switches, we let $F_g=T+1$.

\medskip
For all $(d_1,...,d_T)\in \mathcal{D}$, let $Y_{g,t}(d_1,...,d_T)$ be the potential outcome of group $g$ at period $t$ if $g$ has the treatment path $(d_1,...,d_T)$. This potential outcome model, introduced by \cite{robins1986new}, allows for dynamic effects of lagged treatments on the current outcome, and for anticipation effects of future treatments on the current outcome. However, following most of the literature, we rule out anticipation effects hereafter:\footnote{The notation $Y_{g,t}(d_1,...,d_T)$ also implicitly rules out dependence on treatments that may have occurred before period 1; see Section 1.8 in the web appendix of \cite{de2020difference} for a discussion on this issue.}

\begin{hyp}
	(No Anticipation) For all $g$, $t<T$ and $(d_1,...,d_T)\in \mathcal{D}$, $Y_{g,t}(d_1,...,d_T)$ does not depend on $(d_{t+1},...,d_T)$; we denote it by $Y_{g,t}(d_1,...,d_t)$.
	\label{hyp:no_antic}
\end{hyp}

Assumption \ref{hyp:no_antic} requires that a group's current outcome does not depend on its future treatments. It is plausible when treatment's introduction is hard to anticipate. It is less plausible when treatment's introduction is announced saliently ahead of time. Then, researchers sometimes redefine a $(g,t)$ cell as treated if at period $t$, it has been announced that group $g$ will get treated in the future.

\medskip
The potential outcome notation $Y_{g,t}(d_1,...,d_t)$ implicitly assumes that groups' treatment prior to period one, the first time period in the data, does not affect their outcome, the so-called ``initial conditions'' assumption. When groups might have been exposed to treatment before period one, this assumption is not innocuous, though very few papers have attempted to relax it in the DID literature.

\begin{hyp}
	(i.i.d. groups) The variables $((Y_{g,1}(d_1),...,Y_{g,T}(d_1,...,d_T))_{(d_1,...,d_T)\in \mathcal{D}},\bm{D}_g)_{g\ge 1}$ are i.i.d.
	\label{hyp:iid}
\end{hyp}

Given that groups are identically distributed, we omit the index $g$ hereafter in the absence of ambiguity.


\subsection{Complex designs} 
\label{sub:complex_designs}

We first describe several common designs that are not classical designs or binary and staggered adoption designs.

\paragraph{Non-absorbing binary treatments.}
First, social scientists are often interested in the effect of a non-absorbing binary treatment. For instance, \cite*{burgess2015value} study the effect, in Kenya, of sharing the ethnicity of a country's president, on a district's volume of public expenditures. Districts can enter and leave the treatment (sharing the president's ethnicity) twice over the study period.
An interesting special case is when groups can join and leave treatment once:
\begin{equation}\label{eq:general_binaryoneexit}
D_t=1\{E\geq t\geq F\},
\end{equation}
where both $E$ and $F$ are random. When \eqref{eq:general_binaryoneexit} holds, groups may get treated and leave treatment once, at possibly heterogeneous dates $F$ and $E$.

\paragraph{Absorbing treatments with variation in treatment timing and dose.}
Second, social scientists are often interested in the effect of an absorbing treatment with variation in treatment timing and dose:
\begin{equation}\label{eq:general_stag_nonbinary}
D_t=I \times 1\{t\geq F\},
\end{equation}
where both $I$ and $F$ are random ($I>0$). If \eqref{eq:general_stag_nonbinary} holds, treatment is absorbing but there is variation across groups in the period at which they start receiving the treatment, and in the dose they receive.
For instance, \cite{favara2015credit} study the effect, in the US, of financial deregulations conducted during the 1990s, on the volume of credit and housing prices. Their design almost satisfies \eqref{eq:general_stag_nonbinary}: US states deregulate at heterogeneous times and with heterogeneous intensities. The only difference is that a small number of states deregulate more than once over the study period, so strictly speaking the treatment is not absorbing.

\paragraph{Treatments that vary at baseline.}
Third, social scientists are often interested in the effect of treatments whose intensity varies across groups at all time periods, including at period one:
\begin{equation}\label{eq:general_variation_baseline_treatment}
V(D_1)>0.
\end{equation}
For instance, \cite{gentzkow2011} study the effect, in the US, of the number of newspapers in circulation in a county on turnout in presidential elections in that county. In 1868, the first presidential election used in their analysis, counties' number of newspapers ranges from 0 to 33. Another example is
\cite*{fuest2018higher}, who study the effect, in Germany, of the local business tax rate on wages. In 1993, the first period in their data, municipalities have business tax rates ranging from 10 to 37 percentage points.

\paragraph{Restriction on the design.} We seek to consider as general designs as possible, to include in particular the cases above. Nonetheless, we impose Assumption \ref{hyp:design} below.

\begin{hyp}
\label{hyp:design}
(Designs with some stayers) $\Dr:=\{d\in\mathcal{D}_1:\; V(F|D_{1}=d)>0\}\ne \emptyset$.\footnote{\label{foot:regular_condit_V} In case the distribution of $D_1$ is continuous, $V(F|D_{1}=d)$ is only defined on a set of $d$'s of measure 1. Then, we implicitly assume in what follows that there exists a continuous version of $d\mapsto V(F|D_{1}=d)$.}
\end{hyp}

We thus require that there exists an initial treatment value for which there is heterogeneity in the date at which groups change treatment for the first time. This requirement is natural in difference-in-differences contexts, and there are many applications where it holds. Still, it fails in designs without stayers, where $D_2\ne D_1$ and $F=2$ almost surely, as will for instance be the case if $D_{g,t}$ is the amount of rainfall or the average temperature in location $g$ and year $t$: all locations will experience different precipitations or temperatures in years one and two. This also fails if groups all change treatment for the first time at the same date $t_0$, for instance due to a universal policy affecting them all: $F=t_0$ almost surely. In such cases, if all groups are untreated at period one and receive heterogeneous treatment doses at $t_0$, the design is actually an heterogeneous adoption design and one can then use the estimators considered by \cite*{de2022heterogeneousadoption}.


\subsection{Parallel trends} 
\label{sub:parallel_trends}

We consider two versions of the parallel trends assumption. The first is the classical one. Hereafter, we let $\0_t$ denote the vector of $t$ zeros (similarly, we use below $\1_t$ to denote the vector of $t$ ones).
\begin{hyp}
	(Parallel trends for the never-treated outcome) For all $t\geq 2$, $E\big[Y_{t}(\0_t) -$ $Y_{t-1}(\0_{t-1})|\bm{D}\big]=E\big[Y_{t}(\0_t) -$ $Y_{t-1}(\0_{t-1})\big]$.
	\label{hyp:pt_nevertreated}
\end{hyp}

\begin{hyp}
	(Parallel trends if groups' treatment never changes, conditional on their period-one treatment)
	\label{hyp:PTSQ}
If $D_{1}\in \Dr$, we have, for all $t\ge 2$,
$$E[Y_{t}(D_1,...,D_1) - Y_{t-1}(D_1,...,D_1)|\bm{D}] = E[Y_{t}(D_1,...,D_1) - Y_{t-1}(D_1,...,D_1)| D_{1}].$$
\end{hyp}
$Y_{t}(D_1,...,D_1)$ denotes groups' outcome in the counterfactual where they keep their period-one treatment $D_1$ from period one to $t$, hereafter referred to as their status-quo potential outcome. Assumption \ref{hyp:PTSQ} requires that groups' outcome evolutions in the status-quo counterfactual do not depend on their actual treatment path, once we condition on $D_1$. If all groups are untreated at period 1 and $V(F|D_{1}=0)>0$, we have $\Dr=\{0\}$. Then, Assumption \ref{hyp:PTSQ} is equivalent to Assumption \ref{hyp:pt_nevertreated}. Note that Assumption \ref{hyp:PTSQ} restricts only one potential outcome per group, so Assumption \ref{hyp:PTSQ} alone does not restrict groups' treatment effects.

\medskip
Assumptions \ref{hyp:pt_nevertreated} or \ref{hyp:PTSQ} may be both seen as strict exogeneity conditions. One may worry that non-absorbing designs arise because those that receive the treatment self-select in and out of it, and self-selection makes such strict exogeneity conditions implausible. However,  in most of the aforementioned examples, the multiple treatment changes come from laws that are changed several times or repealed after having been enacted. Therefore, while it is important to document the reasons that led the legislator to further or cancel an initial policy change, parallel-trends assumptions are not by construction less plausible in non-absorbing designs.



\section{Estimation using parallel trends only}
\label{sec:par_trends}

\subsection{Non-normalized event-study effects} 
\label{sub:difference_in_difference_estimators}

\subsubsection{Parameters of interest} \label{ssub:disagg_non_norm}

We first define our parameters of interest, and to this end we introduce additional notation. First, let $S=\sgn(D_F-D_1)$ if $F\le T$, $S=0$ otherwise. Namely, $S=1$ for groups whose treatment increases when it first switches, $S=-1$ for groups whose treatment decreases, and $S=0$ for groups whose treatment never changes. Second, let us define
$$\Dnc_t := \left\{(d_1,...,d_T)\in \mathcal{D}: \;\text{either } \min_{t\ge s>1} d_s \ge d_1 \text{ or } \max_{t\ge s>1} d_s \le d_1\right\}.$$	
In words, $\Dnc_t$ is the set of treatment paths $(d_1,...,d_T)$ without ``crossing'' until $t$: there cannot be treatment values $d_s$ and $d_{s'}$ ($s, s'\le t$) both above and below the initial treatment value ($d_s< d_1<d_{s'}$). Finally, let $\overline{T}_d:=\max\Supp(F-1|D_1=d)$ and $\overline{T}:=\overline{T}_{D_1}$. The random variable $\overline{T}$ represents, for a given group, the last date at which we can find with a positive probability a ``control group'' with the same initial treatment value and whose treatment has not switched yet.

\medskip
Now, for any $\ell \in\{1,...,T-1\}$ for which $P(D_1\in\Dr, F-1+\ell \le \overline{T}, \bm{D}\in\Dnc_{F-1+\ell})>0$, let
$$\AVSQ_\ell=E\left[S\times \left(Y_{F-1+\ell}-Y_{F-1+\ell}(D_1,...,D_1)\right)|D_1\in\Dr, F-1+\ell \le \overline{T}, \bm{D}\in\Dnc_{F-1+\ell}\right].$$
To interpret $\AVSQ_\ell$, let us first suppose that $S\ge 0$ and $\bm{D}\in\Dnc_{F-1+\ell}$ almost surely (a.s.). Then, $\AVSQ_\ell$ is an average expected difference between groups' actual outcome and their counterfactual ``status quo'' outcome if their treatment had always remained equal to their period-one value $D_1$, $\ell$ periods after the first switch occurs. Because of this, we refer to $\AVSQ_\ell$ as an actual-versus-status-quo (AVSQ) event-study effect. The expectation is over  groups for which $D_1\in\Dr, F-1+\ell \le \overline{T}$. Intuitively, and as shown below, these are the groups for which the expected counterfactual outcome can be identified.

\medskip
In a binary and staggered adoption design, $S\ge 0$ and $\bm{D}\in\Dnc_{F-1+\ell}$ a.s., $\Dr=\{0\}$, and thus, $\AVSQ_\ell$ boils down to
$$\text{ATT}_\ell:=E\left[Y_{F-1+\ell}(\0_{F-1},\1_\ell)-Y_{F-1+\ell}(\0_{F-1+\ell})|F-1+\ell \le \overline{T}\right].$$
This is the average effect of having received the treatment for $\ell$ periods of time, among treated groups that become treated early enough to reach $\ell$ periods of exposure to treatment at a time period where there is still an untreated group that can be used as a control (the condition $F-1+\ell \le \overline{T}$). $\text{ATT}_\ell$ is the the event-study effect estimated by \cite{callaway2018}, \cite{borusyak2020revisiting}, and \cite{abraham2018}. Therefore, $\AVSQ_\ell$ generalizes $\text{ATT}_\ell$ to non-binary and/or non-staggered designs.

\medskip
The interpretation of $\AVSQ_\ell$ is more delicate in non-binary and/or non-staggered designs, even if $S\ge 0$ and $\bm{D}\in\Dnc_{F-1+\ell}$ a.s. For instance, if Eq. \eqref{eq:general_binaryoneexit} holds, so that $D_t=1\{E \geq t\geq F\}$ for some random $E$ and $F$, $\AVSQ_\ell$ is a weighted average of
$$E\left[Y_{F-1+\ell}(\0_{F-1},\1_\ell)-Y_{F-1+\ell}(\0_{F-1+\ell})|F-1+\ell \le \min(E, \overline{T})\right]$$
and
$$E\left[Y_{f-1+\ell}(\0_{f-1},\1_{e-(f-1)}, \0_{f-1+\ell-e}) -Y_{f-1+\ell}(\0_{f-1+\ell})|E=e, F=f\right],$$
for all $(e,f)$ satisfying $e< f-1+\ell\le \overline{T}$. The latter expectation is an effect of having been treated for $e-(f-1)$ periods, $f-1+\ell-e$ periods ago. Thus, not only the date at which the effect is evaluated ($F-1+\ell$), but also the number of treatment periods and the recency of the treatment episode varies across groups, complicating the interpretation of $\AVSQ_\ell$. Similarly, with three periods and $\mathcal{D}=\{(0,0,4),(0,1,2),(0,0,0)\}$,
$\AVSQ_1$ is a weighted average of $E(Y_{3}(0,0,4)-Y_{3}(0,0,0)|D_3=4)$ and $E(Y_2(0,1)-Y_2(0,0)|D_2=1)$. Thus, not only the date at which the effect is evaluated but also the magnitude of the treatment increments generating $\AVSQ_\ell$ varies across groups, which again complicates the interpretation of $\AVSQ_\ell$.

\medskip
We introduce the condition $\bm{D}\in\Dnc_{F-1+\ell}$ in the definition of $\AVSQ_\ell$ to ensure that it satisfies the following ``no-sign reversal'' property \citep*{Imbens94,small2017instrumental}:\\
If $(d_1,...,d_t)\ge (d'_1,...,d'_t)$ (where the inequality should be understood component-wise) $ \Rightarrow Y_t(d_1,...,d_t)\ge Y_t(d'_1,...,d'_t)$ a.s., then $\AVSQ_\ell\ge 0$.\\
Suppose we do not condition on $\bm{D}\in\Dnc_{F-1+\ell}$ in $\AVSQ_\ell$ and assume that $\mathcal{D}=\{(1,1,1),(1,2,0)\}$. Then,
\begin{align*}
\AVSQ_2 = & E\left[Y_3(1,2,0)-Y_3(1,1,1)|F=2\right] \\
= & E\left[Y_3(1,2,1)-Y_3(1,1,1)|F=2\right]- E\left[Y_3(1,2,1)-Y_3(1,2,0)|F=2\right].	
\end{align*}
Hence, $\AVSQ_2$ would weight negatively $E\left[Y_3(1,2,1)-Y_3(1,2,0)|F=2\right]$. As a result, we could have $\AVSQ_2<0$ even if increasing the current treatment almost surely increases the outcome.

\medskip
Similarly, it is also to ensure that $\AVSQ_\ell$ satisfies the no-sign reversal that we multiply the difference between the actual and the status-quo outcome by $S$ in its definition. Conditional on $\mathcal{D}^{nc}_{F-1+\ell}$, groups such that $D_F>D_1$, hereafter referred to as ``switchers-in'', are also such that $(D_F,...,D_{F-1+\ell})\ge(D_1,...,D_1)$: the difference between their actual and status-quo outcome is an effect of having exposed to a weakly larger treatment dose for $\ell$ periods. Similarly, groups such that $D_F<D_1$, hereafter referred to as ``switchers-out'', are such that $(D_F,...,D_{F-1+\ell})\le(D_1,...,D_1)$: the difference between their actual and status-quo outcome is an effect of having exposed to a weakly lower dose for $\ell$ periods. Multiplying it by -1 ensures that now, this difference is an effect of having been exposed to a weakly larger dose, which can then be aggregated with switchers-in's AVSQ effects.

\subsubsection{Identification and estimation} 
\label{ssub:identification_and_estimation}

The following theorem shows that under the above assumptions and for suitable $\ell\ge 1$, $\AVSQ_\ell$ is identified by a difference-in-difference estimand. To simplify notation, let us define
\begin{align*}
\Dl := & \bigg\{(d_1,...,d_T)\in \mathcal{D}: d_1\in\Dr, \; f-1+\ell \le \overline{T}_{d_1} \text{ and } (d_1,...,d_T) \in\Dnc_{f-1+\ell},\\
& \quad \text{with } f:=\min\{t>1:d_t\ne d_1\}\bigg\},
\end{align*}
so that $\{D_1\in\Dr, F-1+\ell \le \overline{T},\bm{D}\in\Dnc_{F-1+\ell}\}$ is equivalent to $\bm{D}\in \Dl$.

\begin{thm}\label{thm:ident_AVSQ}
	Suppose that Assumptions \ref{hyp:no_antic}-\ref{hyp:design} and \ref{hyp:PTSQ} hold. Then, provided that $P(\bm{D}\in\Dl)>0$, $\AVSQ_\ell$ is identified. Moreover,
\begin{align}
\AVSQ_\ell = & E\left[S\times(Y_{F-1+\ell}-Y_{F-1} - \Delta_\ell(D_1,F))|\bm{D}\in\Dl\right], \label{eq:ident_AVSQ}	
\end{align}
with $\Delta_\ell(d_1,f) = E[Y_{f-1+\ell}-Y_{f-1}|D_1=d_1,F>f-1+\ell]$. Finally, $P(\bm{D}\in\Dl)>0$ holds for at least $\ell=1$.
\end{thm}

For every switcher, the estimand in Theorem 1 compares, for groups that have first switched $\ell-1$ periods ago, their $F-1$ to $F-1+\ell$ outcome evolution to $\Delta_\ell(D_1,F)$, the average $F-1$ to $F-1+\ell$ outcome evolution of groups with the same $D_1$ as the switcher and whose treatment has not changed yet at period $F-1+\ell$.

\medskip
This theorem implies that when $D_1$ has a finitely supported distribution,\footnote{See \cite{chaisemartin2022continuous} for an analysis of the case when $D_1$ has a continuous distribution.} we can form the following simple estimator of $\AVSQ_\ell$:
$$\widehat{\AVSQ}_\ell = \frac{1}{\# \mathcal{S}_{\ell}} \sum_{g\in \mathcal{S}_{\ell}} S_g\left(Y_{g,F_g-1+\ell}-Y_{g,F_g-1} -\widehat{\Delta}_{g,\ell}\right),$$
where $\# A$ denotes the cardinality of the set $A$,
$$\mathcal{S}_{\ell}:=\big\{g\in\{1,...,G\}: \, F_g-1+\ell\le T,\; \bm{D}_g\in\Dnc_{F_g-1+\ell} \text{ and } \#\mathcal{C}_{g,\ell}>0\big\},$$
$\mathcal{C}_{g,\ell}:=\{g'\in\{1,...,G\}:D_{g',1}=D_{g,1}, F_{g'}>F_g-1+\ell\}$ and
$$\widehat{\Delta}_{g,\ell} := \frac{1}{\# \mathcal{C}_{g,\ell}} \sum_{g'\in \mathcal{C}_{g,\ell}} (Y_{g',F_g-1+\ell}-Y_{g',F_g-1}).$$
When $\mathcal{S}_{\ell}=\emptyset$, we simply let $\widehat{\AVSQ}_\ell =0$. The set $\mathcal{C}_{g,\ell}$ includes the control groups for $g$, namely the groups with the same period-one treatment as switcher $g$ and whose treatment has not changed yet at period $F_g-1+\ell$. Hence, $\mathcal{S}_{\ell}$ denotes the subset of switchers $g$ satisfying the no-crossing condition and for which a control group can be found. The condition $g\in\mathcal{S}_{\ell}$ can therefore be seen as the finite-sample counterpart of $\bm{D}_g\in\Dl$.


\medskip
By slightly adapting the proof of Theorem 1 in \cite{de2020difference}, we can prove that $\widehat{\AVSQ}_\ell$ is consistent and asymptotically normal for $\AVSQ_\ell$, as $G$ tends to infinity, under mild restrictions \citep{de2020difference}.


\subsubsection{Placebo effects} 
\label{ssub:placebo_effects}

An appealing feature of the parallel trend assumption is that it is partly testable. As in classical designs, we can define pre-trend estimands that mimic the estimands identifying the $\AVSQ_\ell$ effects. Specifically, for $\ell\ge 1$, let
$$\AVSQ_{-\ell} = E\left[S\times(Y_{F-1-\ell} - Y_{F-1} - \Delta_{-\ell}(D_1,F)) |\bm{D}\in\Dl, F-1-\ell\ge 1 \right],$$
where $\Delta_{-\ell}(d_1,f) = E[Y_{f-1-\ell}-Y_{f-1}|D_1=d,F>f-1+\ell]$. Under Assumptions \ref{hyp:no_antic}-\ref{hyp:design} and \ref{hyp:PTSQ}, $\AVSQ_{-\ell} = 0$, so under the maintained Assumptions \ref{hyp:no_antic}-\ref{hyp:design}, we can test for Assumption \ref{hyp:PTSQ} by testing this condition. Note that compared to the parameter $\AVSQ_{\ell}$, in $\AVSQ_{-\ell}$ we also condition on $F-1-\ell\ge 1$, to ensure that $Y_{F-1-\ell}$ can be computed. Note that if groups switch early on, we may have $P(\bm{D}\in\Dl, F-1-\ell\ge 1)=0$, even with $\ell=1$; then, $\AVSQ_{-1}$ is undefined and we cannot test Assumption \ref{hyp:PTSQ}. Abstracting from this condition $F-1-\ell\ge 1$, $\AVSQ_{-\ell}$ is computed on the same subpopulation as $\AVSQ_{\ell}$.

\medskip
The estimator of $\AVSQ_{-\ell}$ is similar to that of $\AVSQ_{\ell}$.


\subsubsection{Path-specific effects} 
\label{sub:more_disaggregated_effects}

As mentioned above, $\AVSQ_\ell$ may be delicate to interpret, as it averages the effects of various treatment paths.
Actually, under no-anticipation and parallel trends, we can identify all disaggregated effects of the kind
\begin{equation}
E[Y_t - Y_t(d_1,...,d_1)|\bm{D}=(d_1,...,d_T)],	
	\label{eq:general_TE}
\end{equation}
for all $(d_1,...,d_T)$ such that $d_1\in\Dr$ and $\min\{t:d_t\ne d_1\}\le\max\Supp(F-1|D_1=d_1)$.
If the design is such that the number of treatment paths meeting these conditions is low relative to $G$, then one may be able to precisely estimate such path-specific event-study effects. This may be a feasible approach, for instance, if $D_t=1\{E\ge t\geq F\}$. But in more complicated designs, the number of paths may be too large for this solution to be practical, especially as $\ell$ increases. In such instances, we still recommend that researchers report the treatment paths entering in $\AVSQ_\ell$, as well their distribution: this information may be helpful to interpret $\widehat{\AVSQ}_\ell$.


\subsubsection{Other estimators and identifying strategies} 
\label{sub:other_identifying_strategies}

\paragraph{Difference-in-difference estimators. }

In a binary and staggered design, $\widehat{\AVSQ}_1$ is numerically equal to the $\DID_M$ estimator in \cite{dcDH2020}, and for all $\ell\ge 1$, $\widehat{\AVSQ}_\ell$ is numerically equal to the event-study estimator of the effect of $\ell$ periods of exposure to treatment of \cite{callaway2018}, using the not-yet treated as controls. Outside of binary and staggered designs, when all groups are untreated at period one, $\widehat{\AVSQ}_\ell$ is numerically equal to the estimator obtained by redefining the treatment as an indicator equal to one if group $g$'s treatment has ever changed at $t$, and then computing the event-study estimator of $\ell$ periods of exposure to treatment of \cite{callaway2018} with this binarized and staggerized treatment. This ``binarize and staggerize'' idea has for instance been used by \cite{deryugina2017fiscal} or \cite{krolikowski2018choosing}. When groups' period-one treatment varies, the two estimators are not equal:\footnote{For instance, \cite*{east2023multi} consider designs where groups' period-one treatment varies, and binarize and staggerize the treatment and compute the event-study estimators of \cite{callaway2018}.}
$\widehat{\AVSQ}_\ell$ only compares switchers and not-yet-switchers with the same period-one treatment, whereas the estimator of \cite{callaway2018} applied to this binarized and staggerized treatment compares switchers and non-switchers with different period-one treatments. Then, that estimator relies on a different parallel trend assumption, namely
\begin{equation}
E\left[Y_{g,t}(D_1,...,D_1)-Y_{g,t-1}(D_1,...,D_1)|\bm{D}\right] = E\left[Y_{g,t}(D_1,...,D_1)-Y_{g,t-1}(D_1,...,D_1)\right].	
	\label{eq:strong_par_trends}
\end{equation}
When combined with Assumption \ref{hyp:pt_nevertreated}, one can show that contrary to Assumption \ref{hyp:PTSQ}, \eqref{eq:strong_par_trends} rules out effects of lagged treatments and/or  time-varying treatment effects. We refer to  Section 3.1 in \cite{de2020difference} for a detailed discussion of this issue.

\paragraph{Imputation estimators.}

Our estimator of $\AVSQ_\ell$ consists in estimating the missing counterfactual outcome $Y_{g,F_g-1+\ell}(D_{g,1},...,D_{g,1})$ by $Y_{g,1} + \widehat{\Delta}_g$. Instead, one could consider an imputation estimator, following \cite{borusyak2020revisiting}, \cite{gardnertwo}, and \cite*{liu2021practical}. Their papers focus on the case where $D_1=0$, but when $D_1$ varies we propose to extend their method as follows. For each $d_1\in\Dr$, we regress $Y_{g,t}$ on groups and time fixed effects, on the subsample $\{(g,t):D_{g,t}=d_1,t<F_g\}$. Let $\widehat{\alpha}_g$ and $\widehat{\gamma}_{d_1,t}$ denote the corresponding estimated group and time fixed effects. Then, we impute $Y_{g,F_g-1+\ell}(D_{g,1},...,D_{g,1})$ by $\widehat{\alpha}_g + \widehat{\gamma}_{D_{g,1},F_g-1+\ell}$. Given the asymptotic results on the imputation estimator in \cite{borusyak2020revisiting}  when $D_1=0$, we anticipate that this estimator would also be asymptotically normal as $G$ tends to infinity, provided that for each $d_1$, the number of groups for which $\{(g,t):D_{g,t}=d_1,t<F_g\}$ also goes to infinity.

\medskip
Finally, in principle we could use other identifying restrictions than parallel trends, e.g. those underlying synthetic controls \citep*{abadie2010synthetic}, factor models \citep{xu2017generalized}, or synthetic difference-in-differences \citep*{arkhangelsky2021synthetic}. Those methods also rely on imputing $Y_{g,t}(D_1,...,D_1)$. However, they may be challenging to apply in complex designs, as they require a large number of periods before any treatment change, and a large number of control groups for each set of switchers with a specific value of $D_1$.



\subsection{Normalized event-study effects} 
\label{sub:normalized_event_study_effects}

\subsubsection{Definition of the parameter} 
\label{ssub:definition_of_the_parameter}

Caution must be used when comparing the different $(\AVSQ_\ell)_{\ell\ge 1}$. Not only the elapsed time since the first switch, but also the average increase in received treatment doses (compared to the status-quo) varies with $\ell$. For this reason, we now consider a normalized version of $\AVSQ_\ell$, which allows one to interpret $\AVSQ_\ell$ as a weighted average of marginal effects. To this end, let
$$\AVSQ^D_\ell := E\left[S\sum_{k=0}^{\ell-1} \left(D_{F+k}-D_1\right)|\bm{D}\in\Dl\right].$$
Hence, $\AVSQ^D_\ell$ represents the average change (always counted positively) in total treatment doses between the actual treatment paths and the status quo. This average is taken over the same groups as in $\AVSQ_\ell$. Note that in binary staggered adoption designs, we simply have $\AVSQ^D_\ell=\ell$. In designs satisfying \eqref{eq:general_stag_nonbinary} ($D_t=I \times \ind{t\ge F}$), $\AVSQ^D_\ell=\ell\times E[I]$.

\medskip
Then, the normalized event-study effect we consider is
$$\AVSQ^n_\ell := \frac{\AVSQ_\ell}{\AVSQ^D_\ell}.$$
We define placebo effects as $\AVSQ^n_{-\ell} := \AVSQ_{-\ell}/\AVSQ^D_\ell$ for $\ell\ge 1$. Note that we simply modify the numerator here, since $\AVSQ^D_{-\ell}=0$ by construction.


\subsubsection{Link with effects of specific lags} 
\label{ssub:link_with}

To better interpret $\AVSQ^n_\ell$, let us consider more disaggregated effects, where only one specific lag is modified. For $k\in \{0,...,\ell-1\}$, let
\begin{align}
\text{SL}^\ell_k := \frac{1}{D_{F-1+\ell-k}-D_{1}}\bigg[ & Y_{F-1+\ell}(D_1,...,D_{F-1+\ell-k-1},\underline{D_{F-1+\ell-k}},D_1,...,D_1) \notag \\
& - Y_{F-1+\ell}(D_1,...,D_{F-1+\ell-k-1},\underline{D_{1}},D_1,...,D_1)\bigg]. \label{eq:def_slope}
\end{align}
be the slope of the potential outcome $\ell-1$ periods after the first switch, when the $k$-th lag (the underlined terms) is switched from its status-quo counterfactual value $D_{1}$ to its actual value $D_{F-1+\ell-k}$, whereas all previous treatments are held at their actual values, and all subsequent treatments are held at their status-quo value.\footnote{Thus, when $k=0$, there are actually no $D_1$ terms after the underlined treatment values in \eqref{eq:def_slope} (and the 0-th treatment lag is simply the group's current treatment). We also use in \eqref{eq:def_slope} the convention 0/0=0.}

\medskip
Next, for any $k\in \{0,...,\ell-1\}$, let
$$\omega^\ell_k=\frac{S\times(D_{F-1+\ell-k}-D_{1})}{\AVSQ^D_\ell}.$$
Because $S\times(D_{F-1+\ell-k}-D_{1})\ge 0$ and $\AVSQ^D_\ell>0$, $\omega^\ell_k\ge 0$. Moreover, by definition of $\AVSQ^D_\ell$, $E\left[\sum_{k=0}^{\ell-1} \omega^\ell_k |\bm{D}\in\Dl\right]=1$. Finally, note that
$$\sum_{k=0}^{\ell-1} \omega^\ell_k \text{SL}^\ell_k = S\times(Y_{F-1+\ell}(D_1,...,D_{F-1},...,D_{F-1+\ell}) - Y_{F-1+\ell}(D_1,...,D_1)).$$
As a result, we obtain that $\AVSQ^n_{\ell}$ is the expectation of a weighted average of $(\text{SL}^\ell_k)_{k=0,...,\ell-1}$, with positive weights:

\begin{thm}\label{thm:norm_delta_ell}
For every $\ell\ge 1$ for which $P(\bm{D}\in\Dl)>0$,
\begin{equation}
\AVSQ^n_{\ell}=E\left[\sum_{k=0}^{\ell-1} \omega^\ell_k \text{SL}^\ell_k \,\big|\, \bm{D}\in\Dl\right].	
	\label{eq:decomp_AVSQ_n}
\end{equation}
Moreover, $E\left[\sum_{k=0}^{\ell-1} \omega^\ell_k |\bm{D}\in\Dl\right]=1$ and $\omega^\ell_k\geq 0$.
\end{thm}

The decomposition \eqref{eq:decomp_AVSQ_n} simplifies much in designs where groups' treatment can change at most once ($D_{g,t}=D_{g,F_g}$ for all $t\ge F_g$). Then, $\omega^\ell_k=1/\ell$ for all $k=0,...,\ell-1$. As a result, $\AVSQ^n_1$ is an effect of the current treatment on the outcome, $\AVSQ^n_{2}$ is a weighted average of the effect of the current treatment and of the first treatment lag on the outcome with weights 1/2, and so on. When groups' treatment can change more than once, estimating $k \mapsto \Omega^\ell_k:= E[\omega^{\ell}_k|\bm{D}\in\Dl]$ helps documenting which lags contribute the most to $\AVSQ^n_{\ell}$.

\medskip
The shape of $\ell\mapsto \AVSQ^n_{\ell}$ may provide insights as to whether the outcome is more affected by recent or by old treatments. To see this, suppose that $\ell \mapsto \Omega^\ell_k$ is decreasing for all $\ell>k$ and assume a linear model $Y_t(d_1,...,d_t)=\mu_t+\sum_{k=0}^t \delta_k d_{t-k}$, in which case $\text{SL}^\ell_k = \delta_k$ for all $k\in\{0,...,\ell-1\}$ and $\ell\ge 1$ for which $P(\bm{D}\in\Dl)>0$. Then, $\AVSQ^n_{\ell}=\sum_{k=0}^{\ell-1} \Omega_k^\ell \delta_k$. In this case, one can show that if $k\mapsto \delta_k$ is decreasing, $\ell\mapsto \AVSQ^n_{\ell}$ is decreasing as well. Hence, a decreasing $\ell\mapsto \AVSQ^n_{\ell}$ provides suggestive evidence that potential outcomes are more affected by recent than by past treatments. 	


\subsubsection{Identification and estimation} 
\label{ssub:identification_and_estimation_n}

It follows directly from the definition of $\AVSQ^n_\ell$ and the results on $\AVSQ_\ell$ that $\AVSQ^n_\ell$ can be identified by
$$\AVSQ^n_\ell = \frac{E\left[S\times(Y_{F-1+\ell}-Y_{F-1} - \Delta_\ell(D_1,F))|\bm{D}\in\Dl\right]}{E\left[S\sum_{k=0}^{\ell-1} \left(D_{F+k}-D_1\right)|\bm{D}\in\Dl\right]}.$$
Like that of $\AVSQ_\ell$, its plug-in estimator is asymptotically normal as $G$ tends to infinity, under mild restrictions.



\subsection{Cost-benefit analysis}\label{subsec:general_HRDID_dyn_cost-benefit}

We show here that the actual-versus-status-quo event-study effects can be used to perform a cost-benefit analysis. Let us  suppose that groups always switch to a weakly larger treatment than their period-one treatment:
\begin{equation}\label{eq:general_HRDID_dyn_increasing_design}
D_t \ge D_1 \quad \forall t\ge 2 \quad \text{a.s.} 
\end{equation}
We impose \eqref{eq:general_HRDID_dyn_increasing_design} to reduce the notational burden. When it fails, one can just conduct separate cost-benefit analyses for groups such that $S=1$ and groups such that $S=-1$.

\medskip
Now, let us consider the parameter
$$\ACE=\frac{\sum_{\ell=1}^{T-1}P(F-1+\ell\le \overline{T}) \AVSQ_\ell}{\sum_{\ell=1}^{T-1}P(F-1+\ell\le \overline{T})E[D_{F-1+\ell}-D_1|F-1+\ell\le\overline{T}]}.$$

As explained in \cite{de2020difference}, $\ACE$ corresponds to an average cumulative effect per unit of treatment, whence its name. By definition of $\AVSQ_\ell$ and \eqref{eq:general_HRDID_dyn_increasing_design}, we can rewrite the $\ACE$ as
$$\ACE:=\frac{E\left[\sum_{\ell=1}^{\overline{T}-(F-1)} Y_{F-1+\ell}-Y_{F-1+\ell}(D_1,...,D_1)\right]}{ E\left[\sum_{\ell=1}^{\overline{T}-(F-1)} (D_{F-1+\ell}-D_1)\right]},$$
using the convention that the sums are 0 if $\overline{T}\le F-1$. Now, let us take the perspective of a planner, seeking to conduct a cost-benefit analysis comparing groups' actual treatments $\bm{D}$ to the counterfactual ``status-quo'' scenario where they would have always kept their period-one treatment. Assume that the outcome is a measure of output, such as agricultural yields or wages, expressed in monetary units. Assume also that the treatment is costly, with a cost linear in dose, and known to the analyst. Then, let $C_\ell\geq 0$ denote the (random) cost of administering one treatment dose in a given group, $\ell-1$ periods after this group first switches. Assuming that the planner's discount factor is equal to $1$, groups' actual treatments are beneficial in monetary terms relative to the status quo, up to period $\overline{T}$, if and only if
$$E\left[\sum_{\ell=1}^{\overline{T}-(F-1)} Y_{F-1+\ell}-Y_{F-1+\ell}(D_1,...,D_1)\right] \ge  E\left[\sum_{\ell=1}^{\overline{T}-(F-1)} C_\ell(D_{F-1+\ell}-D_1)\right].$$
Equivalently, $\ACE \ge c$, where
$$c:=\frac{E\left[\sum_{\ell=1}^{\overline{T}-(F-1)} C_\ell(D_{F-1+\ell}-D_1)\right]}{E\left[\sum_{\ell=1}^{\overline{T}-(F-1)} (D_{F-1+\ell}-D_1)\right]}$$
is the average treatment cost, across all the incremental treatment doses
received with respect to the status-quo counterfactual. Then, comparing the $\ACE$ to $c$ is sufficient to evaluate if changing groups' treatments from their initial treatments to $\bm{D}$ was beneficial.

\medskip
Two additional remarks on the $\ACE$ are in order. First, in binary and staggered designs, if no group is treated at period one and there are never-treated groups ($\overline{T}=T$), we have
$$\ACE=\sum_{t=1}^T \frac{P(D_t=1)}{\sum_{t'=1}^T P(D_{t'}=1)} E\left[Y_t(\0_{F-1},\1_{t-(F-1)}) - Y_t(\0_t)|D_t=1\right].$$
The expectation on the right-hand side may be seen as the ATT at period $t$, so the right-hand side may be seen as the global ATT, over the $T$ periods. Hence, the $\ACE$ generalizes the ATT to non-binary and/or non-staggered designs.

\medskip
Second, it follows directly from the results on $\AVSQ_\ell$ that we can consistently estimate the $\ACE$ by
$$\widehat{\ACE} = \frac{\sum_{\ell=1}^{T-1}\widehat{P}(F-1+\ell\le \overline{T}) \widehat{\AVSQ}_\ell}{ \sum_{\ell=1}^{T-1}\widehat{P}(F-1+\ell\le \overline{T})\widehat{E}[D_{F-1+\ell}-D_1|F-1+\ell\le\overline{T}]}.$$

\subsection{Testing the null that lagged treatments do not affect the outcome}
\label{par:xdh_a_joint_test}

While allowing for dynamic effects is appealing, doing so reduces the sample that can be used in the
estimation, which may come with high costs in terms of external validity and statistical precision \citep*{chaisemartin2022continuous}. We show here that parameters closely related to the $\AVSQ_\ell$ can be used to test the null that there are no dynamic effects, namely:
\begin{hyp}\label{hyp:no_dyn}
	(No Dynamic Effects) For all $g$ and $(d_1,...,d_t)\in \mathcal{D}$, $Y_{g,t}(d_1,...,d_t)=Y_{g,t}(d_t)$.
	\end{hyp}
Not rejecting those tests may suggest we could use estimators ruling out dynamic effects \citep[see, e.g.,][]{chaisemartin2022continuous}, which may lead to external-validity and precision gains with respect to the estimators discussed in this paper.

\subsubsection{Tests when some switchers eventually revert to their initial treatment value} 
\label{ssub:a_test_with_units_switching_out_of_treatments}

In designs with a binary treatment and where some groups leave the treatment after having been previously treated, \cite{liu2021practical} propose a test of Assumption \ref{hyp:no_dyn} under the standard parallel-trends condition (Assumption \ref{hyp:pt_nevertreated}), which amounts to estimating the average treatment effect across previously treated groups, at time periods where those groups have left the treatment. Under Assumption \ref{hyp:no_dyn}, this average treatment effect should be equal to zero. Their test is implemented by the \st{fect} Stata \citep*{fectStata} and R \citep*{fectR} commands.

\medskip
We can extend this test to non-binary treatments, under our alternative parallel-trends condition (Assumption \ref{hyp:PTSQ}). Specifically, let us define
$$\AVSQ^o_\ell:= E\left[S\left(Y_{F-1+\ell}- Y_{F-1+\ell}(D_1,...,D_1)\right)|D_{F-1+\ell}=D_1, \bm{D}\in\Dl\right],$$
for all $\ell\ge 2$ for which $P(D_{F-1+\ell}=D_1, \bm{D}\in\Dl)>0$ (note that $\AVSQ^o_1$ is undefined since by construction $D_F\ne D_1$). Under Assumption \ref{hyp:no_dyn}, $Y_{F-1+\ell}=Y_{F-1+\ell}(D_{F-1+\ell})$, and thus,
$$\AVSQ^o_\ell=0.$$ 
This can be tested easily. In particular, the Stata command \st{did\_multiplegt\_dyn} estimates the $(\AVSQ_\ell)_{\ell\ge 1}$. To estimate $\AVSQ^o_\ell$ for a specific $\ell\ge 2$, it suffices to use the command on the subsample of $(g,t)$ cells such that $D_{g,F_g-1+\ell}=D_{g,1}$ or $t<F_g$.


\subsubsection{Tests without switchers reverting to their initial treatment value} 
\label{ssub:a_test_with_absorbing_treatments}

A disadvantage of the previous tests is that they can only be used when some switchers eventually revert to their initial treatment value. This does not occur, for instance, with absorbing treatments, e.g. in staggered adoption designs. In such cases, testing static effects is more challenging because basically, treatment effects remain unconstrained even if effects are static. As explained earlier, under no-anticipation and parallel trends, we can identify all disaggregated effects of the kind
\begin{equation}
E[Y_t - Y_t(d_1,...,d_1)|\bm{D}=(d_1,...,d_T)],	
	\label{eq:general_TE_st}
\end{equation}
for all $(d_1,...,d_T)$ such that $d_1\in\Dr$ and $\min\{t:d_t\ne d_1\}\le\max\Supp(F-1|D_1=d_1)$. Under  Assumption \ref{hyp:no_dyn}, $Y_t - Y_t(d_1,...,d_1)= Y_t(d_t)-Y_t(d_1)$, so the effect in the previous display reduces to
\begin{equation*}
E[Y_t(d_t)-Y_t(d_1)|\bm{D}=(d_1,...,d_T)],	
\end{equation*}
but that effect may still depend on $\bm{D}$ in an unrestricted way, if $P(D_{F-1+\ell}= D_1|F-1+\ell\le \overline{T})=0$. Therefore, no-anticipation, parallel trends and Assumption \ref{hyp:no_dyn} do not have any testable implication.

\medskip
To make progress, we add an additional assumption, and propose a joint test of no-anticipation, parallel trends, Assumption \ref{hyp:no_dyn}, and that additional assumption. Specifically, on top of Assumption \ref{hyp:no_dyn}, we also assume the following:
\begin{equation}
E[Y_t(d)- Y_t(d')|\bm{D}] = \beta(d, d', \bm{D}), \quad \forall (d,d')\in\mathcal{D}_t^2.	
	\label{eq:restr_TE_for_static}
\end{equation}
Condition \eqref{eq:restr_TE_for_static} allows for treatment effects to depend on the treatment path in an unrestricted way, but requires that they are time-invariant. In particular, if $D_t=D_{t+1}$,
$$E[Y_{t+1}- Y_{t+1}(D_1)|\bm{D}] = \beta(D_{t+1},D_1,\bm{D})= E[Y_t- Y_t(D_1)|\bm{D}].$$
This point is key to obtain our test, based on Proposition \ref{prop:for_static_test}.

\begin{prop}
	Suppose that \eqref{eq:restr_TE_for_static} and Assumptions \ref{hyp:no_antic}-\ref{hyp:design}, \ref{hyp:PTSQ}, and \ref{hyp:no_dyn}  hold, and $L\ge 2$ satisfies $P(D_F=...=D_{F-1+L},\bm{D}\in\mathcal{D}^L)>0$. Then,
$$\AVSQ^{\text{bal}}_\ell := E[S\times(Y_{F-1+\ell} - Y_{F-1+\ell}(D_1,...,D_1))|D_F=...=D_{F+L},\bm{D}\in\mathcal{D}^L],$$
defined for $\ell\in\{1,...,L\}$, does not depend on $\ell$.
\label{prop:for_static_test}
\end{prop}
Proposition \ref{prop:for_static_test} implies that under Assumptions \ref{hyp:no_antic}-\ref{hyp:design} and \ref{hyp:PTSQ}, we can construct a simple test of both Assumption \ref{hyp:no_dyn}
and \eqref{eq:restr_TE_for_static}: we just need to estimate $(\AVSQ^{\text{bal}}_\ell)_{\ell \in \{1,...,L\}}$, and test that all effects are equal. For that purpose, one can use the \st{did\_multiplegt\_dyn} command on the subsample of $(g,t)$ cells such that $D_{g,F_g}=D_{g,F_g+1}=...=D_{g,F_g+L}$ or $t<F_g$, specifying the \st{same\_switchers} and \st{effects\_equal} options.

\medskip
Compared to the test of \cite{liu2021practical}, the test based on $(\AVSQ^{\text{bal}}_\ell)_{\ell=1,...,L}$ is a joint test, rather than a test of Assumption \ref{hyp:no_dyn} alone. This makes a rejection of this test more difficult to interpret. On the other hand, this test is feasible even if $P(D_{F-1+\ell}= D_1|F-1+\ell\le \overline{T})=0$, whereas the previous one is not. Even if $P(D_{F-1+\ell}= D_1|F-1+\ell\le \overline{T})>0$, there may be few groups reverting back to their initial treatment, in which case the test of \cite{liu2021practical} may have lower power than the test based on $(\AVSQ^{\text{bal}}_\ell)_{\ell=1,...,L}$.



\section{Imposing additional restrictions} 
\label{sec:learning}

We have seen so far that in complex designs with non-absorbing and/or non-binary treatment, we can still identify reduced-form parameters that may be policy relevant, to evaluate the policies that were conducted with respect to what would have happened without any policy. Identification is obtained under no-anticipation and parallel-trends assumptions, whose plausibility can be assessed via pre-trends tests, as in standard DID estimation.

\medskip
However, this approach also has limitations. While normalized event-study effects estimate weighted averages of marginal effects of the current and lagged treatments on the outcome, they cannot be used to separately estimate those marginal effects. Moreover, the cost-benefit analysis considered above allows one to compare the actual policy with the status-quo, but cannot evaluate alternative policies, or determine the optimal policy.

\medskip
We address these issues by considering the following additional restriction:

\begin{hyp} (Random coefficients distributed-lag linear model)
There exist a random vector $B:=(\beta_{0},\dots,\beta_{K})'$ such that
$$Y_{t}(d_1,...,d_t) = Y_{t}(\0_t)+ \sum_{k=0}^{\min(K,t-1)} \beta_{k} d_{t-k},$$
where $K<T-1$ is known. Moreover, $E[\|B\|^2]<\infty$.
\label{hyp:RC_model}
\end{hyp}
Assumption \ref{hyp:RC_model} allows for heterogeneous treatment effects across groups, and it does not impose any restriction on the dependence between the random coefficients $B$ and the treatment path $\bm{D}$.
On the other hand, Assumption \ref{hyp:RC_model} also imposes some restrictions. First, it assumes that the effects of the current treatment and its lags do not vary over time. This condition is strong, but imposing it seems unavoidable if one wants to use past policy changes to do an ex-ante evaluation of future policies: with time-varying effects, the effects of past policy changes cannot be used to infer the effect of future policies. Second, it assumes that the analyst knows the number of lags up to which past treatments still affect the current outcome. Third, it assumes that the effects of the current treatment and its lags are linear, and that they do not complement or substitute each other.

\medskip
We first study below what the usual distributed-lag TWFE regressions identify under no anticipation, parallel trends and Assumption \ref{hyp:RC_model}. Because of the possible correlation between $B$ and $\bm{D}$, such regressions may fail to identify a causal effect. We then consider how to identify expectations of $B$. Finally, we consider relaxations of some of the restrictions imposed by Assumption \ref{hyp:RC_model}.

\subsection{Distributed-lag two-way fixed effects regressions}

In general designs, a commonly-used estimator of dynamic treatment effects, for instance discussed in Equation (5.2.6) of \cite{angrist2009mostly}, is the distributed-lag TWFE estimator. For $k \in \{0,...,K\}$, let $\widehat{\beta}_{k}$ denote the coefficient on $D_{g,t-k}$ in a regression of $Y_{g,t}$ on group and period FEs and $\left(D_{g,t-k}\right)_{k \in \{0,...,K\}}$, in the subsample such that $t\geq K+1$:
\begin{equation}\label{eq:general_twfe_lags}
Y_{g,t}=\sum_{g'=1}^G\widehat{\alpha}_{g'}1\{g=g'\}+\sum_{t'=1}^T\widehat{\gamma}_{t'}1\{t=t'\}+\sum_{k=0}^K\widehat{\beta}_{k}D_{g,t-k}+\hat{\epsilon}_{g,t}.
\end{equation}
In practice, researchers may slightly augment or modify \eqref{eq:general_twfe_lags}. They may include treatment leads in the regression, to test Assumptions \ref{hyp:no_antic} and \ref{hyp:PTSQ}. They may define the lagged treatments as equal to 0 at time periods when they are not observed, and estimate the regression in the full sample. They may also estimate the regression in first difference and without group fixed effects. Finally, they may include control variables. Results similar to Theorem \ref{thm:general_distributedlag} below apply to all those variations on \eqref{eq:general_twfe_lags}.

\medskip
Before stating this result, we introduce additional notation. For $g=1,...,G$, let $\eta^k_{g,t}$ denote the sample residual of cell $(g,t)$ ($t\ge K+1$) in the regression of $D_{g,t-k}$ on group and time fixed effects and $(D_{g,t-k'})_{k'\ne k}$. Then, for any $(k,k')\in\{0,...,K\}^2$, let
$$W^{k,k'}_g=\frac{\sum_{t\ge K+1} \eta^k_{g,t} D_{g,t-k'}}{(1/G) \sum_{g'=1}^G \sum_{t\ge K+1} \eta^k_{g't}D_{g',t-k}}.$$
This random variable depends on $G$ but we omit the dependence to simplify notation. Since the initial sample is i.i.d., the distribution of $W^{k,k'}_g$ does not depend on $g$, so we omit this dependence below.

\begin{thm}\label{thm:general_distributedlag}
Suppose that Assumptions \ref{hyp:no_antic}-\ref{hyp:pt_nevertreated}  and \ref{hyp:RC_model} hold. Then, for all $k \in \{0,...,K\}$,
\begin{equation}
E\left[\widehat{\beta}_{k}\right]=E\bigg[W^{k,k} \beta_k + \sum_{\substack{k'=0 \\ k'\neq k}}^K~W^{k,k'} \beta_{k'}\bigg].
	\label{eq:TWFE_weights}
\end{equation}
Moreover, $E[W^{k,k}]=1$ and $E[W^{k,k'}]=0$ for all $k'\ne k$. Finally, if $\Cov(W^{k,k'},\beta_{k'})=0$ for all $k' \in \{0,...,K\}$, $E\left[\widehat{\beta}_{k}\right]=E\left[\beta_{k}\right]$.
\end{thm}

The proof of Theorem \ref{thm:general_distributedlag} is very similar to that of Theorem 2 in \cite{de2020two}, but we include it in the appendix for completeness. It essentially extends Proposition 3 in \cite{abraham2018} to non-binary or non-staggered designs. Theorem \ref{thm:general_distributedlag} shows that even under Assumption \ref{hyp:RC_model}, which is restrictive, the distributed-lag TWFE estimator $\widehat{\beta}_{k}$ does not identify $E\left[\beta_{k}\right]$ in general. Rather, it estimates the sum of $K+1$ terms. The first term is a weighted average of $\beta_{k}$, where the weight variable has expectation one but may be negative. Then, if treatment effects are heterogeneous between groups, we may not have $E[W^{k,k} \beta_k]= E[\beta_k]$. If $P(W^k<0)>0$, we may even have a sign reversal, namely $P(\beta_{k}>0)=1$ yet $E\left[\widehat{\beta}_{k}\right]<0$. Now, \eqref{eq:TWFE_weights} also includes a sum of $K$ terms. These terms include $\beta_{k'}$, multiplied by another ``weight'' $W^{k,k'}$, which has expectation 0. Thus, $\widehat{\beta}_{k}$, which is supposed to estimate the effect of the $k$th treatment lag, is actually contaminated by the effects of other treatment lags. If $V(\beta_{k'})=0$, these contamination terms disappear as the expectation of the contamination weights is equal to zero. But if $V(\beta_{k'})>0$, the contamination terms may differ from zero and can also lead to sign reversal.

\medskip
Importantly, the weights $(W_g^{k,k'})_{g=1,...,G}$ can be computed, and this computation can help diagnosing the robustness of the distributed-lag TWFE regression to heterogeneous treatment effects. For instance, a useful diagnostic is to look at the variance of the weights $(W_g^{k,k})_{g=1,...,G}$ around their expectation, equal to one. A large variance indicates that the regression strongly upweights the $\beta_k$ of some groups and strongly downweights or even weights negatively the $\beta_k$ of other groups, and if the weights are correlated with groups' treatment effects, this could bias $\widehat{\beta}_{k}$ far away from $E[\beta_k]$.

\medskip
Finally, note that Theorem \ref{thm:general_distributedlag} is derived under Assumption \ref{hyp:pt_nevertreated}, a parallel-trends assumption on the never-treated outcome, while event-study estimands in the previous section relied on Assumption \ref{hyp:PTSQ}, a parallel-trends assumption on the status-quo outcome. In Section 1 of their Web Appendix, \cite{de2020two} provide an example where a TWFE regression is actually ``less robust'' under a parallel-trends assumption in the spirit of Assumption \ref{hyp:PTSQ} than under a parallel-trends assumption in the spirit of Assumption \ref{hyp:pt_nevertreated}.


\subsection{Heterogeneous distributed-lag regressions} 
\label{sub:robust_distributed_lag_regressions}

\subsubsection{Motivation and setup} 
\label{ssub:setup}

Under Assumption \ref{hyp:RC_model}, the slope $\text{SL}^\ell_k$ defined in \eqref{eq:def_slope} reduces to $\beta_k$. Then, Theorem \ref{thm:norm_delta_ell} shows that $\AVSQ^n_{\ell}$ identifies a weighted average of expectations of the $\beta_k$s, for $k$ ranging from $0$ to $\ell-1$. Actually, as discussed in a working paper version of \cite{de2020difference}, under Assumption \ref{hyp:RC_model} one can leverage DID estimators to separately estimate the expectation of each  $\beta_k$. First, one estimates group-specific regressions of
$$(Y_{g,F_g-1+\ell}-Y_{g,F_g-1} -\widehat{\Delta}_{g,\ell})_{\ell\in \{1,...,\overline{T}-(F_g-1)\}},$$
the building-block DID estimators in the previous section,
on $(D_{g,F_g-1+\ell}-D_{g,1})_{\ell\in \{1,...,\overline{T}-(F_g-1)\}}$ and its lags. Then, one averages  those group-specific regression coefficients across switchers \citep[see Section 4 of][]{de2020differencev5}. In this paper, we will not use DIDs to estimate the expectation of each  $\beta_k$. Instead, we will propose estimators that may be used in more general designs, namely in designs that may not have stayers.

\medskip
First, we show that when combined with  the standard parallel trends assumption (Assumption \ref{hyp:pt_nevertreated}), Assumption \ref{hyp:RC_model} leads to a particular case of the random coefficients model considered by \cite{chamberlain1992efficiency} and \cite{arellano2012identifying}. To see this, first note that Assumption \ref{hyp:pt_nevertreated} is equivalent to
$$\Delta Y_{t}(\0_t) = \gamma_t + \eps_{t},\quad E[\eps_{t}|\bm{D}]=0,$$
where $\Delta$ is the first-difference operator. Next, let $\Delta\bm{Y}=(\Delta Y_{K+2},\dots,\Delta Y_{T})$, $\Gamma=(\gamma_{K+2},\dots,\gamma_T)'$, $\bm{\eps} = (\eps_{K+1},\dots,\eps_{T})'$ and
$$\bm{M}= \begin{pmatrix} \Delta D_{K+2} & \dots & \Delta D_{2} \\ \Delta D_{K+3} & \dots & \Delta D_{3} \\
\vdots & \vdots & \vdots \\
\Delta D_{T} & \dots & \Delta D_{T-K}  	
\end{pmatrix}.$$
Then,
\begin{equation}
\Delta \bm{Y} =  \Gamma+ \bm{M} B + \bm{\eps}, \quad E[\bm{\eps} | \bm{M}]=0.	
	\label{eq:RC_model}
\end{equation}
We now introduce an identifying assumption on the design, discussed in details below. For any matrix $A$, let $A^+$ denote its Moore-Penrose inverse, $I$ denote the identity matrix (without specifying its size in the absence of ambiguity) and let $\Pi(A):=I-AA^+$ be the orthogonal projector on the orthocomplement of the image of $A$.

\begin{hyp} (Condition on the design)
$E[\Pi(\bm{M})]$ is invertible.
\label{hyp:design_RC}
\end{hyp}


\subsubsection{Identification} 
\label{ssub:identification_rc}

We consider identification of average effects of the kind $E[\beta_k|A]$ for some subpopulation $A$ specified below. First, because $\Pi(\bm{M})\bm{M}=0$, it follows from \eqref{eq:RC_model} that under Assumption \ref{hyp:design_RC}, $\Gamma$ is identified by
\begin{equation}
\Gamma = E[\Pi(\bm{M})]^{-1} E[\Pi(\bm{M})\Delta \bm{Y}].	
	\label{eq:ident_Gamma}
\end{equation}
Then, let $e_k$ denote the $k+1$-th canonical vector in $\R^{K+1}$ (e.g., $e_0=(1,0,...,0)'$). Informally, we can identify $\overline{\beta}_k:=E[\beta_k |e_k\in \text{Im}(\bm{M}')]$, provided that $P(e_k\in \text{Im}(\bm{M}'))>0$, by
\begin{equation}
\overline{\beta}_k = E\left[e'_k\bm{M}^+(\Delta Y -\Gamma)|e_k\in \text{Im}(\bm{M}')\right].
    \label{eq:ident_B_informal}
\end{equation}
Heuristically, \eqref{eq:ident_B_informal} follows from the fact that when $e_k\in \text{Im}(\bm{M}')$, $e'_k\bm{M}^+\bm{M}=e'_k$ and thus $e'_k\bm{M}^+(\Delta Y -\Gamma)=\beta_k+e_k'\bm{M}^+\bm{\eps}$. This identification argument is informal, though, because as discussed in \cite{graham2012identification} and \cite{chaisemartin2022continuous} in related setups, the expectation on the right-hand side of \eqref{eq:ident_B_informal} may fail to exist. Namely, we may have
\begin{equation}
E\left[|e'_k\bm{M}^+(\Delta Y -\Gamma)| \, \big|\, e_k\in \text{Im}(\bm{M}')\right] = \infty,
	\label{eq:infinite_exp}
\end{equation}
due to groups for which $\|\bm{M}'{}^+e_k\|$ is arbitrary large. In Appendix \ref{sub:infinite_exp}, we show that indeed, \eqref{eq:infinite_exp} holds if $T=2$, $K=0$, $\Delta D_2$ has a positive density around 0 and a weak regularity condition holds on $\eps_2$. Nevertheless, the following theorem shows that (i) we can still identify  $\overline{\beta}_k$ by a suitable truncation and a continuity argument; (ii) the concern above does not apply if $D_t$ is finitely supported for all $t$.

\begin{thm}\label{thm:ident_RC}
	Suppose that Assumptions \ref{hyp:no_antic}, \ref{hyp:iid}, \ref{hyp:pt_nevertreated}, \ref{hyp:RC_model} and \ref{hyp:design_RC} hold and $E[\Delta D_t^2+\Delta Y_t(\0_t)^2]<\infty$ for all $t\ge 2$. Then $\Gamma$ is identified by \eqref{eq:ident_Gamma}. Moreover, if $P(e_k\in \text{Im}(\bm{M}')>0$ for $k\in\{0,...,K\}$, $\overline{\beta}_k$ is identified by
\begin{equation}
\overline{\beta}_k = \lim_{C\uparrow \infty}
E\left[e'_k\bm{M}^+(\Delta Y -\Gamma)|e_k\in \text{Im}(\bm{M}'), \|\bm{M}'{}^+e_k\|\le C\right]
	\label{eq:ident_B}
\end{equation}
Finally, if $D_t$ is finitely supported for all $t=1,...,T$, the right-hand side of \eqref{eq:ident_B_informal} is well-defined and $\overline{\beta}_k$ is identified by \eqref{eq:ident_B_informal}.
\end{thm}

This theorem is similar to Proposition 1 in \cite{arellano2012identifying}, with a few differences. First, Model \eqref{eq:RC_model} is a particular case of their random coefficient model. Second, we consider the identification of averages of each component of $B$, rather than the identification of an average of the full vector $B$. This may be important in practice, because in the former case, we just need to restrict the estimation to groups for which $e_k\in \text{Im}(\bm{M}')$, whereas in the latter case we need to restrict the estimation to groups for which $\bm{M}'\bm{M}$ is invertible, and such groups may fail to exist (this is the case for instance if $T<2(K+1)$). The third difference with Proposition 1 in \cite{arellano2012identifying}
is that we highlight the fact that identification can be achieved without trimming when $D_t$ is finitely supported, but may require trimming otherwise.

\medskip
With respect to the identification results in Section \ref{sec:par_trends}, Theorem \ref{thm:ident_RC} applies to a broader class of designs. In particular, expectations of $(\beta_k)_{k\in \{0,...,\ell-1\}}$ may be identified in designs that do not have stayers for long enough for $\AVSQ_\ell$ to be identified, especially when $T>2(K+1)$, namely when the $(T-(K+1))\times (K+1)$ matrix $\bm{M}$ has more rows than columns.\footnote{Intuitively, this helps for Assumption \ref{hyp:design_RC} because then, the image of $\bm{M}$ has dimension at most $K+1$, and thus the matrix $\Pi(\bm{M})$ has rank at least $T-2(K+1)>0$.} For instance, assume that $K=1$, $T=5$ and $\mathcal{D}$, the support of $\bm{D}$, satisfies $\mathcal{D}=\{(0,1,0,1,0), (0,0,1,1,0)\}$. Then, one can show that Assumption \ref{hyp:design_RC} holds, and $\overline{\beta}_0$ and $\overline{\beta}_1$ are identified. On the other hand, with such a design, $\AVSQ_2$ is not identified because $P(D_1=D_2=D_3)=0$. More generally, when $T>2(K+1)$ we can expect $E[\Pi(\bm{M})]$ to be invertible without requiring $P(D_1=...=D_{K+2})>0$, which is necessary to identify $\AVSQ_{K+1}$. In fact, simulations suggest that Assumption \ref{hyp:design_RC} holds if $\bm{D}$ is continuous with respect to the Lebesgue measure on $\R^T$, a case where there is no stayer as $D_2\ne D_1$ and $F=2$ a.s, thus implying that Assumption \ref{hyp:design} fails.

\medskip
When $T \le 2(K+1)$, $E[\Pi(\bm{M})]$ may not be invertible without groups that keep the same treatment value over consecutive periods, so averages of the random coefficients $(\beta_k)_{k\in \{0,...,\ell-1\}}$ are identified under conditions similar to those under which $\AVSQ_\ell$ is identified. For instance, Assumption \ref{hyp:design_RC} fails if $\bm{D}$ is continuous with respect to the Lebesgue measure on $\R^T$. 
Then, $P(\bm{M} \text{ is invertible})=1$ and $\Pi(\bm{M})=0$ a.s. Even with a binary treatment, if $K=1$ and $T=4$, one can show that a necessary condition for invertibility is
\begin{equation}
\min\left(P(D_1=D_2=D_3), P(D_2=D_3=D_4)\right)>0,	
	\label{eq:cond_stayers}
\end{equation}
a condition under which $\AVSQ_2$ is typically identified.

\subsubsection{Estimation} 
\label{ssub:estimation_rc}

Theorem \ref{thm:ident_RC} suggests the following simple, plug-in estimator of $\Gamma$:
$$\widehat{\Gamma} = \left(\frac{1}{G}\sum_{g=1}^G \Pi(\bm{\bm{M}_g})\right)^{-1}\frac{1}{G}\sum_{g=1}^G \Pi(\bm{\bm{M}_g}) \Delta\bm{Y}_g.$$
This estimator is root-$G$ consistent and asymptotically normal, even when $T=2(K+1)$, as soon as Assumption \ref{hyp:design_RC} holds. This contrasts with Proposition 1.1 in \cite{graham2012identification}, who consider a closely related random coefficients model: their result indicates that the common parameters ($\Gamma$ in our context, $\bm{\delta}_0$ in theirs) cannot be estimated at the standard, root-$G$ rate when the number of time periods is equal to the number of regressors ($T=2(K+1)$ in our context). An important difference with their setup is that we restrict the random coefficients to be time-invariant. This allows us to identify $\Gamma$ using observations for which $\bm{M}$ is singular.

\medskip
Following Theorem \ref{thm:ident_RC}, we discuss separately the estimation of $\overline{\beta}_k$ depending on whether $D_t$ is finitely supported or not. 

\paragraph{Finitely supported treatment.} 
\label{par:finitely_supported_treatment}

In this case, we can consider the following simple plug-in estimator:
\begin{equation}
\widehat{\overline{\beta}}_k = \frac{1}{\# \mathcal{S}^{\text{rc}}}\sum_{g\in \mathcal{S}^{\text{rc}}} e_k' \bm{M}_g^+(\Delta \bm{Y}_g - \widehat{\Gamma}),	
	\label{eq:def_Bhat}
\end{equation}
where $\mathcal{S}^{\text{rc}}:=\{g\in\{1,...,G\}: e_k\in\text{Im}(\bm{M}_g')\}$. Like that of $\Gamma$, this estimator is root-$G$ consistent and asymptotically normal under Assumption \ref{hyp:no_antic}-\ref{hyp:pt_nevertreated} and \ref{hyp:RC_model}-\ref{hyp:design_RC}. Its asymptotic variance can be obtained by seeing $(\widehat{\Gamma}, \widehat{\overline{B}})$ as a joint GMM estimator, see, e.g., \cite{newey1984method}.


\paragraph{Not finitely supported treatment.} 
\label{par:other_cases}

When $D_t$ is not finitely supported, the situation is more delicate. In particular, if \eqref{eq:infinite_exp} holds, we have to rely on \eqref{eq:ident_B} for identification. Then, we expect root-$G$ estimation of $\overline{B}$ to be impossible. Equation \eqref{eq:ident_B} suggests the following estimator, which is close to an estimator of \cite{graham2012identification}:
$$\widehat{\overline{\beta}}_k  = \frac{1}{\# \mathcal{S}_C^{\text{rc}}}\sum_{g\in \mathcal{S}_C^{\text{rc}}} e_k' \bm{M}_g^+(\Delta \bm{Y}_g - \widehat{\Gamma}),$$
for some $C>0$ and where $\mathcal{S}_C^{\text{rc}}:=\{g\in\{1,...,G\}: e_k\in\text{Im}(\bm{M}_g'), \,\|\bm{M}_g'{}^+e_k\| \le C\}$. When choosing $C$, we face a usual trade-off between bias, which is small when $C$ is large, and variance, which is small when $C$ is small because we trim groups for which $e_k' \bm{M}_g^+(\Delta \bm{Y}_g - \widehat{\Gamma})$ has a large variance (due to $\|\bm{M}_g'{}^+e_k\|$ being large). Choosing an appropriate $C$ may, however, be delicate in practice.





\subsection{Further results}

\subsubsection{Allowing for dynamic effects up to any treatment lag}

In this section, we replace Assumption \ref{hyp:RC_model} by the following condition:
\begin{hyp} (Random coefficients distributed-lag linear model, with effects up to any lag)\label{hyp:RC_model2}
There exists a random vector $B:=(\beta_{0},...,\beta_{T-1})'$ such that $E[\|B\|^2]<\infty$ and
$$Y_{t}\left(d_1,...,d_t\right) = Y_{t}(\0_t)+ \sum_{k=0}^{t-1} \beta_{k} d_{t-k}.$$
\end{hyp}
Then, we have
$$\Delta Y = \Gamma + \bm{M} B + \bm{\eps},$$
where now $\Delta Y=(\Delta Y_2,...,\Delta Y_T)'$, $\bm{\eps}=(\eps_2,...,\eps_T)'$ $\Gamma = (\gamma_2,...,\gamma_T)'$ and
$$\bm{M} = \begin{pmatrix}
    \Delta D_2 & 0 & & \dots & 0 \\
    \Delta D_3 & \Delta D_2 & 0 & \dots & 0 \\
    \vdots & &  & \ddots & \vdots \\
    \Delta D_T & & \dots & & \Delta D_2
\end{pmatrix}.$$
Then, identification and estimation under Assumption \ref{hyp:RC_model2} proceeds as in Sections \ref{ssub:identification_rc} and \ref{ssub:estimation_rc}. Remark that $\bm{M}$ is now a $(T-1)\times (T-1)$ matrix. Then, one can show that Assumption \ref{hyp:design_RC} holds if and only if there are some ``never switchers'', namely $P(D_1=...=D_T)>0$.

\medskip
Assumption \ref{hyp:RC_model2} allows all lagged treatments up to period one to affect the outcome. Thus, on that dimension it is less restrictive than Assumption \ref{hyp:RC_model}, which only allows the first $K$ lags to have an effect. At the same time, note that under Assumption \ref{hyp:RC_model2},
\begin{equation}\label{eq:obsYunderRC_model2}
Y_{t}= Y_{t}(\0_t)+ \sum_{k=0}^{t-1} \beta_{k} D_{t-k}:
\end{equation}
the number of lags allowed to affect the outcome depends on $t$, the period when the outcome is measured. This reflects an implicit assumption embedded in our potential outcome notation, namely that treatments prior to period one have no effect on groups' outcomes from period one to $T$. If all lags can affect the outcome, namely if instead of \eqref{eq:obsYunderRC_model2} we have that
$$Y_{t}= Y_{t}(\0_t)+ \sum_{k=0}^{+\infty} \beta_{k} D_{t-k},$$
then the estimators introduced in this section can still be used if groups' treatments do not change before period 1:
\begin{equation}\label{eq:nochangebeforeperiodone}
\forall t\leq 0, ~D_t=D_1 \text{ almost surely.}
\end{equation}
\eqref{eq:nochangebeforeperiodone} is a strong condition. Still, it holds by construction when the treatment does not exist before period one. When the treatment is absorbing, it also holds by construction in the subsample of untreated groups at period one. If treatments prior to period one have an effect on groups' outcomes and groups' treatments changed before period one, then the estimators introduced in this section may be misleading. Instead, the estimators proposed in Section \ref{ssub:estimation_rc} can be used in such cases, at the expense of ruling out effects of lagged treatments beyond the $K$th lag, and dropping the outcomes $(Y_t)_{t=1,...,K}$ from the estimation.

\subsubsection{Allowing for interaction effects}

In this section, we replace Assumption \ref{hyp:RC_model} by the following condition, which allows for interaction effects between treatment lags:
\begin{hyp} (Random coefficients distributed-lag linear model with interactions)
There exist random variables $(\beta_{k,k'})_{0\le k\le k'\le K}$ such that
$$Y_{t}(d_1,...,d_t) = Y_{t}(\0_t)+ \sum_{k=0}^K \left[\beta_{k,k} + \sum_{k'=k+1}^K \beta_{k,k'}d_{t-k'} \right] d_{t-k},$$
where $K<T-1$ is known. Moreover, $E[\beta_{k,k'}^2]<\infty$ for all $0\le k\le k'\le K$.
\label{hyp:RC_model_inter}
\end{hyp}
Then, \eqref{eq:RC_model} still holds, except that now the matrix $\bm{M}$ is of dimension $(T-K+1)\times \frac{(K+1)(K+2)}{2}$, and includes columns of the kind $\Delta (D_{t-k} D_{t-k'})$ for some fixed $(k,k')$ and where $t$ varies over the column. As a result, identification may fail even in some canonical designs. For instance, with a binary and absorbing treatment, one cannot tease out the effects of $D_{t-1}D_t$ and $D_{t-1}$: as no treated group goes back to being untreated,  $D_{t-1}D_t=D_{t-1}$.\footnote{Formally, the identifying condition corresponding to $P(e_k\in\text{Im}(\bm{M}'))>0$ in Theorem \ref{thm:ident_RC} does not hold.}
On the other hand, identification may still be achieved with non-absorbing treatments. For instance, if $K=1$, $T=5$ and $\mathcal{D}=\{(0,0,0,0,0),(0,1,1,0,0)\}$, we can identify averages of $\beta_{0,0}$, $\beta_{0,1}$, and $\beta_{1,1}$  on groups for which $\bm{D}=(0,1,1,0,0)$.

\section{Application to \cite{gentzkow2011}} 
\label{sub:application_to_gentzkow_et_al}

In this application, we revisit the effect of newspapers on voters' participation in elections, studied by \cite{gentzkow2011}. They use a US panel data set at the county $\times$  presidential-election level, with 1,195 counties and from the 1868 to the 1928 election. They seek to test a conjecture in \cite{de1850democratie}, that newspapers
encourage citizens to participate more in democratic institutions. For that purpose, they let $Y_{g,t}$ denote the turnout rate in county $g$ and the presidential election that took place in year $t$, they let $D_{g,t}$ denote the number of newspapers circulating in county $g$ and year $t$, and they run a static first-difference regression of $\Delta Y_{g,t}$ on $\Delta D_{g,t}$, and state-year fixed effects. Hereafter, we consider possibly dynamic effects. We first estimate the AVSQ event-study effects, under parallel trends only. Then, we estimate a standard distributed-lag TWFE regressions, before estimating the random coefficients distributed-lag linear model.

\begin{figure}[H]
\begin{center}
    \caption{Distribution of $D_1$ and the number of switches in $D_t$}
    \includegraphics[trim=5mm 10mm 0mm 0mm, width=0.9\textwidth, height=7cm, clip=true]{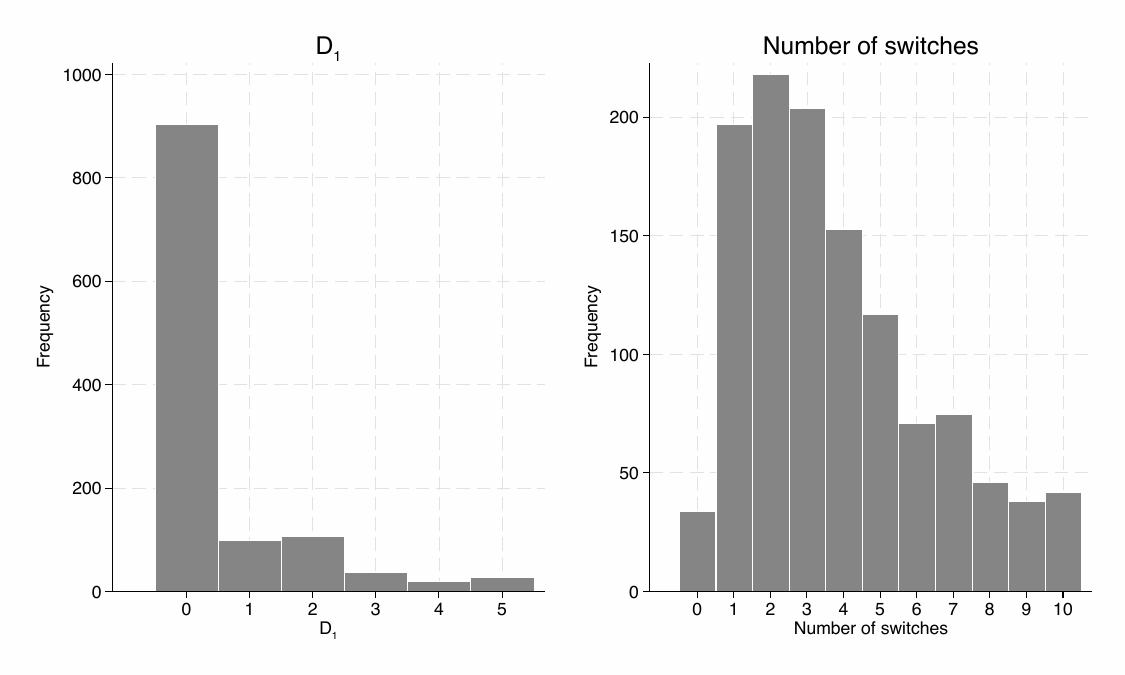}
\label{fig:stat_des}
\end{center}
\footnotesize
\vspace{-0.2cm}
Notes: the last bins (5 and 10) actually correspond to 5 or more and 10 or more.
\end{figure}

\medskip
First, note that the design of this application is truly complex: the number of newspapers is a non-binary treatment, it can increase or decrease over time, counties can experience several changes in their number of newspapers, at different points in time. The right panel of Figure \ref{fig:stat_des} displays the distribution of the number of switches in $D_t$. It shows that over the sixteen years of presidential elections, counties experiencing eight or more changes are not uncommon. Actually, only 34 counties (2.9\%) never experience any change in their number of newspapers. For 90.9\% of the 1,161 remaining counties, $S_g=1$, namely they experience an increase in the number of newspapers at first switch. However, 78.7\% of counties experience at least one decrease in their number of newspapers. Still, the no-crossing condition holds for 93.6\% of the counties. Finally, the left panel of Figure \ref{fig:stat_des} displays the distribution of $D_1$, which also shows significant heterogeneity. Assumption \ref{hyp:design} is easily met, with $\Dr=\{0,1,...,7\}$.

\medskip

\subsection{AVSQ event-study effects} 
\label{sub:application}

\subsubsection{Non-normalized event-study effects} 
\label{ssub:non_normalized_event_study_effects}

Estimates of $\AVSQ_\ell$ and pre-trends estimates are shown in Figure \ref{fig:Gentzkow_nonnormalized} below.
Being exposed to a weakly larger number of newspapers for one electoral cycle increases turnout by 1.44 percentage point, and the effect is statistically significant (s.e.=0.43 percentage point). That effect can be estimated for 1,119 out of the 1,195 counties in the data: 34 counties never experience a change in their number of newspapers, and 42 counties that do experience a change cannot be matched with a not-yet-switcher with the same number of newspapers at baseline. Being exposed to a weakly larger number of newspapers for two, three, and four electoral cycles also significantly increases turnout. Effects increase with exposure length, but one cannot reject the null that all effects are equal (p-value=0.40). As $\ell$ increases, effects mechanically apply to fewer and fewer counties, but the effect after four electoral cycles still applies to 917 counties.

\medskip
Pre-trend estimates are small and individually and jointly insignificant. However, their confidence intervals are quite large.
While the first pre-trend estimator applies to 906 of the 1,119 counties for which $\widehat{\AVSQ}_{1}$ is estimated, the fourth pre-trend estimator only applies to 447 of the 917 counties for which $\widehat{\AVSQ}_{4}$ is estimated. The confidence interval of $\widehat{\AVSQ}_{-4}$ is already quite large, but that of $\widehat{\AVSQ}_{-5}$ is substantially larger, so we have very little power to detect differential trends over more than five election cycles. This is why we only report four placebo and four event-study estimators.

\medskip
\begin{figure}[H]
    \begin{center}
    \caption{Non-normalized DID estimates of effects of newspapers on turnout}
    \includegraphics[trim=3mm 0mm 0mm 10mm,scale=0.9,clip=true]{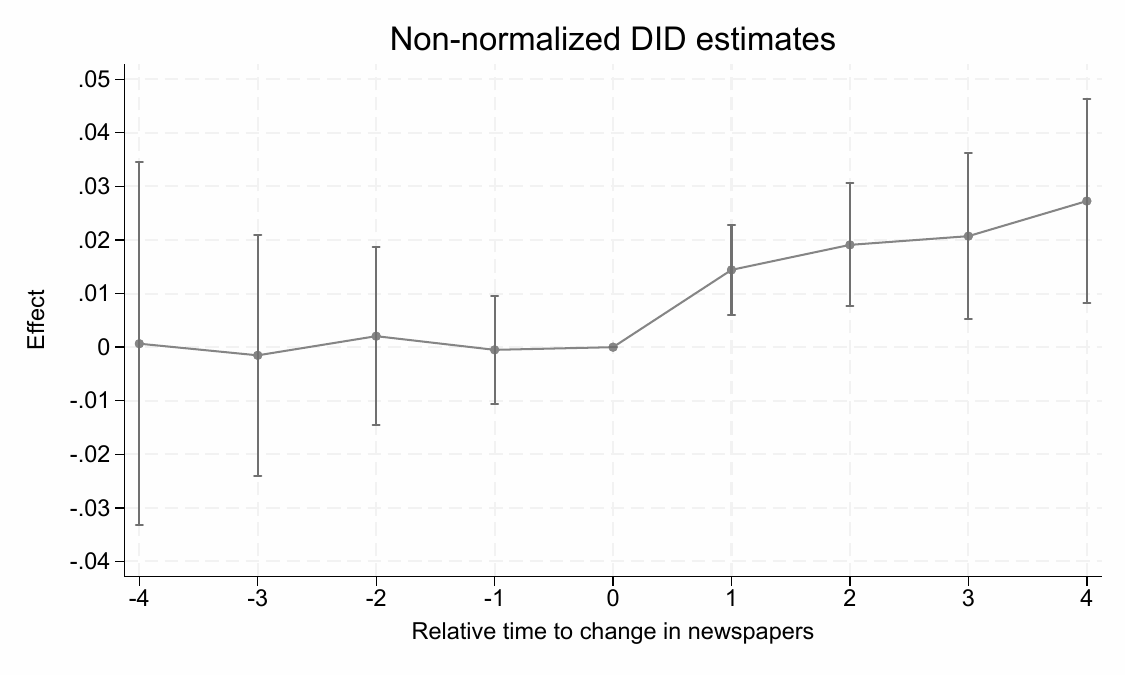}
    \label{fig:Gentzkow_nonnormalized}
    \end{center}
\footnotesize
\vspace{-1cm}
Notes: we computed $\ell \mapsto \widehat{\AVSQ}_{\ell}$ 
using the \st{did\_multiplegt\_dyn} Stata command. Vertical lines: 95\% confidence intervals, based on standard errors clustered at the county level.
\end{figure}

\medskip
To better interpret the $\AVSQ_\ell$, we report the underlying most common treatment paths. For AVSQ$_1$, the three most common effects are effects of having one versus zero newspapers (64\% of the cases), two versus zero newspapers (12\% of the cases) and two versus one newspapers (5\% of the cases). For AVSQ$_2$, the three most common effects are effects of having $D_F=D_{F+1}=1$ instead of $D_F=D_{F+1}=0$ (32\% of the cases), $D_F=1, D_{F+1}=0$ instead of $D_F=D_{F+1}=0$ (18\% of the cases), and $D_F=1, D_{F+1}=2$ instead of $D_F=D_{F+1}=0$ (12\% of the cases). Finally, for AVSQ$_4$, the three most common effects are effects of having
$D_F=...=D_{F+3}=1$ instead of $D_F=...=D_{F+3}=0$ (15\% of the cases), $D_F=1, D_{F+1}=D_{F+2}=D_{F+3}=0$ instead of $D_F=...=D_{F+3}=0$ (14\% of the cases) and $D_F=1, D_{F+1}=D_{F+2}=D_{F+3}=2$ instead of $D_F=...=D_{F+3}=0$ (5\% of the cases). As $\ell$ increases, AVSQ$_\ell$ averages the effects of more and more heterogeneous paths across groups, and the three most common paths account for a smaller fraction of all the effects averaged in AVSQ$_{\ell}$. For most paths, too few groups have that path to estimate reasonably precisely a path-specific effect.


\subsubsection{Normalized event-study effects} 
\label{ssub:normalized_event_study_effects}

Normalized event-study and pre-trends estimates are shown in Figure \ref{fig:Gentzkow_normalized} below. Normalized event-study estimates are decreasing with $\ell$, but one cannot reject the null that all effects are equal (p-value=0.17). $\widehat{\Omega}^1_0=1$: the first event-study estimate is an effect of contemporaneous newspapers on turnout. $\widehat{\Omega}^2_0=0.48$ and $\widehat{\Omega}^2_1=0.52$: the second normalized event-study estimate is a weighted average of the effects of contemporaneous newspapers and of the first lag of newspapers on turnout, with approximately equal weights. $\widehat{\Omega}^3_0=0.35$, $\widehat{\Omega}^3_1=0.31$, and $\widehat{\Omega}^3_2=0.33$: the third normalized event-study estimate is a weighted average of the effects of contemporaneous newspapers and of the first and second lag of newspapers, with approximately equal weights. Finally, $\widehat{\Omega}^4_0=0.28$, $\widehat{\Omega}^4_1=0.26$, $\widehat{\Omega}^4_2=0.23$, and $\widehat{\Omega}^4_3=0.24$: the fourth normalized event-study estimate is a weighted average of the effects of contemporaneous newspapers and of the first, second, and third lag of newspapers, again with approximately equal weights. Then, the fact that normalized event-study estimates are decreasing with $\ell$ may suggest that lagged newspapers have a smaller effect on turnout than contemporaneous newspapers.

\medskip
\begin{figure}[H]
    \begin{center}
    \caption{Normalized DID estimates of effects of newspapers on turnout}
    \includegraphics[trim=3mm 0mm 0mm 10mm,scale=0.9,clip=true]{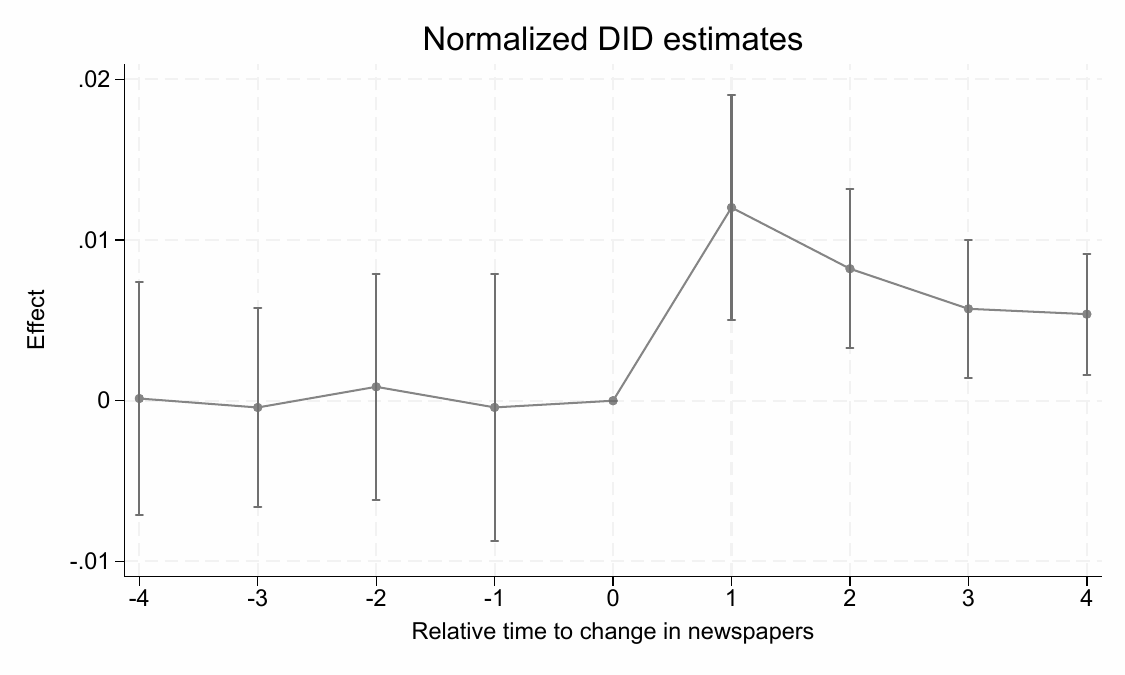}
    \label{fig:Gentzkow_normalized}
    \end{center}
\footnotesize
\vspace{-1cm}
Notes: we computed $\ell \mapsto \widehat{\AVSQ}^n_{\ell}$ 
using the \st{did\_multiplegt\_dyn} Stata command. Vertical lines: 95\% confidence intervals, based on standard errors clustered at the county level.
\end{figure}


\subsubsection{Testing the null that lagged treatments do not affect the outcome} 
\label{ssub:testing_static_effects}

Given that the treatment may revert to its initial value, our first test is applicable here. The results are displayed in Table \ref{tab:test_results}. We cannot reject the null of static effects, though the tests do not have much power, given the low number of switchers which eventually revert to their initial treatment, for which $\AVSQ_\ell^o$ can be estimated. We also conduct the joint test based on Proposition \ref{prop:for_static_test} above. When considering $L=2$, we can estimate $\AVSQ^{\text{bal}}_\ell$ using 512 counties. In that subsample, the estimates of $\AVSQ^{\text{bal}}_1$ and $\AVSQ^{\text{bal}}_2$ are close and not significantly different (p-value=0.83). Again, this suggests that lagged newspapers do not affect turnout.

\medskip
\begin{table}[H]
\begin{center}
\caption{Tests of static effects based on estimates of $\AVSQ_\ell^o$}

\begin{tabular}{l|cccc}
\toprule
$\ell$ & $\widehat{\AVSQ}_\ell\!\!{}^o$ & Std. err. & p-value & N. obs. \\
\midrule
2 & 0.0189 & 0.0151 & 0.212 & 266 \\
3 & 0.0209 & 0.0188 & 0.267 & 261 \\
4 & 0.0132 & 0.0195 & 0.497 & 241 \\
5 & 0.0558 & 0.0321 & 0.083 & 187 \\
\bottomrule
\end{tabular}
\label{tab:test_results}
\end{center}
{\footnotesize Notes: ``N. obs.'' indicates the number of counties to which $\widehat{\AVSQ}_\ell\!\!{}^o$ applies.  Standard errors clustered at the county level.}
\end{table}


\subsection{Standard distributed-lag regressions} 
\label{ssub:application_to_gentzkow_et_al}

Next, we consider a distributed-lag TWFE estimator with just one lag. We obtain $\widehat{\beta}_0=-0.0008$ (s.e.$=0.0014$) and $\widehat{\beta}_1=0.0050$ (s.e.$=0.0015$). Hence, according to this regression, increasing the current number of newspapers insignificantly reduces turnout by 0.08 percentage points. On the other hand, increasing its first lag significantly increases turnout by 0.5 percentage points. These results may look surprising: the normalized event-study effects suggested the opposite, namely that the current number of newspapers would have a larger impact than the number of newspapers four years before. The tests above also do not reject the hypothesis that only the current number of newspapers affects turnout.

\begin{figure}[H]
	\caption{Distribution of $W^{0,0}$ and $W^{0,1}$}
	\begin{minipage}{0.48\textwidth}
		\centering
		\includegraphics[scale=0.45]{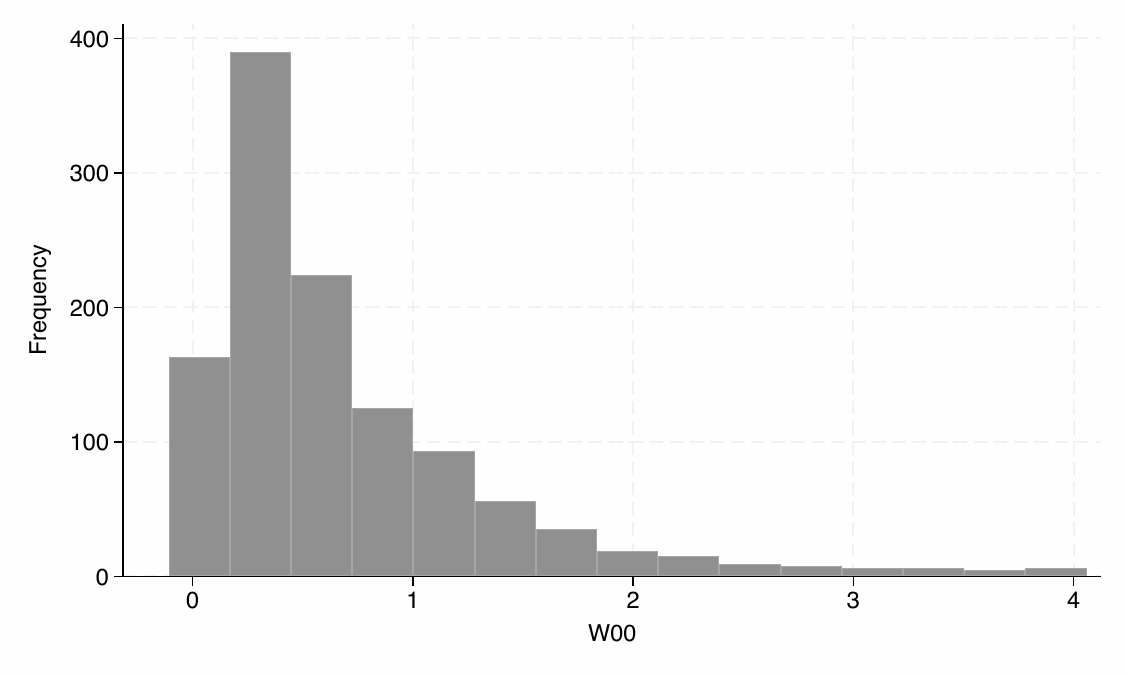}
	\end{minipage}
	\begin{minipage}{0.48\textwidth}
		\centering
		\includegraphics[scale=0.45]{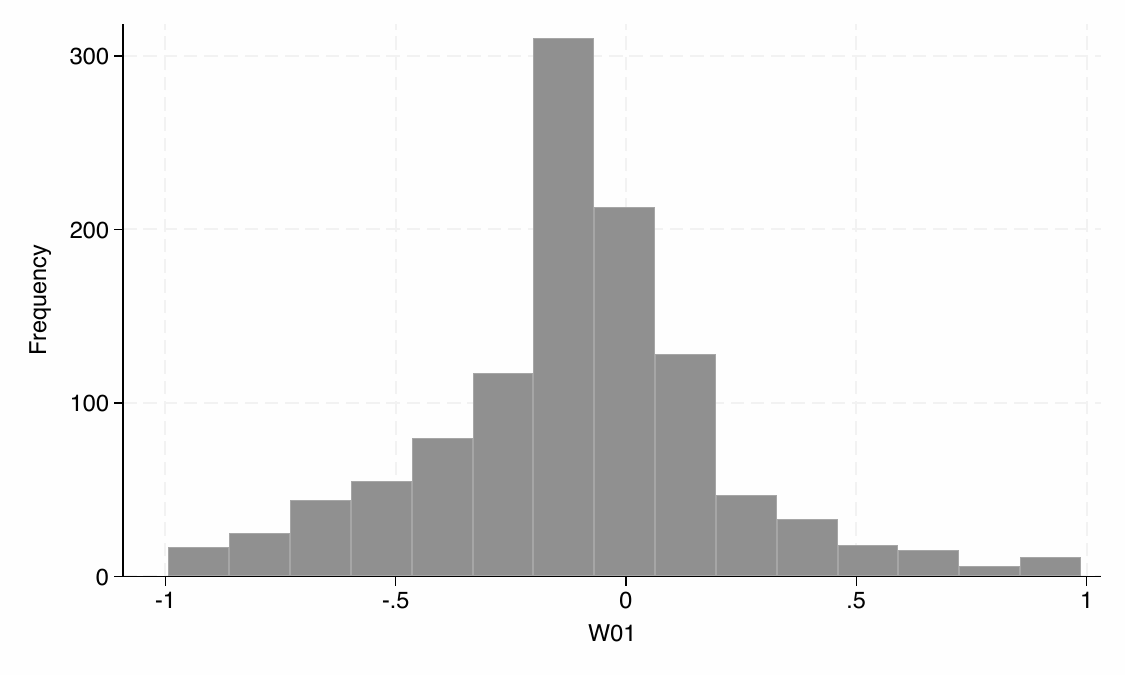}
	\end{minipage}
	\label{fig:hist_weights}
\end{figure}

\medskip
However, conclusions drawn from the distributed-lag TWFE estimator are unwarranted if treatment effects are heterogeneous and correlated with the weights. To assess whether this could be the case, we compute the weights in the decomposition \eqref{eq:TWFE_weights}. Their distribution is plotted in Figure \ref{fig:hist_weights}. Recall that $\widehat{\beta}_0$ estimates the sum of two terms. The first term is a weighted average of 1,195 county-specific effects of current newspapers. Even if most  of the weights $(W_g^{0,0})_g$ are non-negative (1,186 of the 1,195 counties), their standard deviation is large (3.19), with a first quartile equal to 0.22, a 99th percentile equal to 8.8 and a maximum as large as 87.8. This implies that the first term upweights very substantially the effects of a small number of counties, whose effects could well differ from the average effect. The ``weights'' $(W_g^{0,1})_g$, which are centered, also have a large standard deviation (2.09), and vary between -4.65 and 12: contamination of $\widehat{\beta}_0$ by effects of the lagged number of newspapers is substantial.

\medskip
We obtain similar results for $\widehat{\beta}_1$. The standard deviation of the $(W_g^{1,1})_g$ is also large (2.42), with a first quartile equal to 0.16, a 99th percentile equal to 11.8 and a maximum of 26.1. Finally, the ``weights'' $(W_g^{1,0})_g$ also have a large standard deviation (1.39), and vary between -7.1 and 15.8.

\medskip


\subsection{Heterogeneous distributed-lag regressions} 
\label{ssub:heterogeneous_distributed_lag_regressions}

Next, we turn to the estimation of heterogeneous distributed-lag regressions, following Section \ref{sub:robust_distributed_lag_regressions} above. We consider both $K=1$, as above, and $K=2$. The results are displayed in Table \ref{tab:estim_RC_model} below. Note that in both cases, the averages in $\widehat{\overline{\beta}}_k$ apply to a large part of the sample: even $\widehat{\overline{\beta}}_2$ applies to 1,037 counties, namely 90.4\% of the sample. The estimators' standard errors are around 50\% larger than those of the distributed lag TWFE estimates, but they remain relatively precise. Finally, the results are much more in line with the event-study estimates above, and point towards a positive effect of the current number of newspapers, and little effects of the lagged number of newspapers.

\medskip
\begin{table}[H]
\begin{center}
\caption{Estimates of the average effect of each lag}

\begin{tabular}{l|cc|ccc}
\toprule
Model & \multicolumn{2}{c|}{$K=1$} & \multicolumn{3}{c}{$K=2$} \\
Parameter & $\overline{\beta}_0$ & $\overline{\beta}_1$ &
$\overline{\beta}_0$ & $\overline{\beta}_1$ & $\overline{\beta}_2$ \\
\midrule
Estimate & 0.0045 &-0.0008& 0.0042 &-0.0014 &-0.0016\\
Std. err. & 0.0021 &0.0021 & 0.0021 &0.0024 &0.0022\\
N.obs & 1,093 &1,090 & 1,052 & 1,049 & 1,037 \\
\bottomrule
\end{tabular}
\label{tab:estim_RC_model}
\end{center}
{\footnotesize Notes: the two models are estimated on the 1,147 counties for which we have data on consecutive elections and no missing data on $D$. ``N. obs'' denotes the number of counties on which we average effects. Standard errors obtained by bootstrapping counties.}
\end{table}

\doublespacing
\appendix

\section{Appendix} 
\label{sec:proofs}

\subsection{Proof of Theorem \ref{thm:ident_AVSQ}} 
\label{sub:theorem_ref_thm_ident_avsq}

Let us fix $\ell\ge 1$ satisfying $P(\bm{D}\in\Dl)>0$. First, we prove \eqref{eq:ident_AVSQ}. By definition of $F$, we have
$$\Delta_\ell(d_1,f)=E[Y_{f-1+\ell}(D_1,...,D_1)-Y_{f-1}(D_1,...,D_1)|D_1=d_1,F>f-1+\ell].$$
Second, by the parallel-trends condition (Assumption \ref{hyp:PTSQ}), we also have
$$\Delta_\ell(d_1,f)=E[Y_{f-1+\ell}(D_1,...,D_1)-Y_{f-1}(D_1,...,D_1)|D_1=d_1,F=f,S, \bm{D}\in\Dl],$$
since $F$ and $S$ are functions of $\bm{D}$. Hence, $\Delta_\ell(d_1,f)$ is the counterfactual evolution of the outcome under the status-quo between $t=f-1$ and $t=f-1+\ell$, for groups initially at $d_1$ and that switch at $t=f$. Then, by the law of iterated expectations,
\begin{align*}
	& E\left[S(Y_{F-1+\ell}-Y_{F-1} - \Delta_\ell(D_1,F))|\bm{D}\in\Dl\right] \\
	= & E\left[S\left(Y_{F-1+\ell}-Y_{F-1} - E[Y_{F-1+\ell}(D_1,...,D_1)-Y_{F-1}(D_1,...,D_1)|D_1,F,S, \bm{D}\in\Dl]\right)|\bm{D}\in\Dl\right]\\
	= & E\left[S\left(Y_{F-1+\ell}-E[Y_{F-1}|D_1,F,S, \bm{D}\in\Dl] - E[Y_{F-1+\ell}(D_1,...,D_1) \right.\right.\\
	& \quad \left.\left.  -Y_{F-1}(D_1,...,D_1)|D_1,F,S, \bm{D}\in\Dl]\right)|\bm{D}\in\Dl\right]\\
	= & E\left[S\left(Y_{F-1+\ell}- E[Y_{F-1+\ell}(D_1,...,D_1)|D_1,F,S, \bm{D}\in\Dl]\right)|\bm{D}\in\Dl\right] \\
	= & \AVSQ_\ell.
\end{align*}
Next, we prove that there exists $\ell\ge 1$ such that $P(\bm{D}\in\Dl)>0$. We prove the result for $\ell=1$. By definition of $\Dnc_t$, the condition $\bm{D}\in\Dnc_F$ automatically holds. Then,
\begin{align*}
	P(\bm{D}\in\mathcal{D}^1) & = P(D_1\in\Dr, F\le \overline{T}) \\
	& = \sum_{d_1\in\Dr} P(D_1=d_1) P(F< \max\Supp(F|D_1=d_1)|D_1=d_1) \\
	& >0.
\end{align*}
The last inequality follows since $\Dr\ne \emptyset$ by Assumption \ref{hyp:design} and for all $d_1\in\Dr$, $P(D_1=d_1)>0$ and $P(F< \max\Supp(F|D_1=d_1)|D_1=d_1)>0$ (as $V(F|D_1=d_1)>0$).


\subsection{Proof of Proposition \ref{prop:for_static_test}} 
\label{sub:proof_of_proposition_eqref_prop_for_static_test}

For all $\ell=1,..., L$, we have
\begin{align*}
	\AVSQ^{\text{bal}}_\ell & = E[S\times E(Y_{F-1+\ell} - Y_{F-1+\ell}(D_1)|\bm{D}) |D_F=...=D_{F+L},\bm{D}\in\mathcal{D}^L] \\
	& = E[S\times \beta(D_{F-1+\ell},D_1,\bm{D}) |D_F=...=D_{F+L},\bm{D}\in\mathcal{D}^L] \\
	& = E[S\times \beta(D_F,D_1,\bm{D}) |D_F=...=D_{F+L},\bm{D}\in\mathcal{D}^L],	
\end{align*}
which does not depend on $\ell$.


\subsection{Proof of Theorem \ref{thm:general_distributedlag}} 
\label{sub:proof_of_theorem_ref}

By the Frisch-Waugh-Lovell theorem,
$$\widehat{\beta}_k = \frac{\sum_{g=1}^G \sum_{t\ge K+1} \eta_{g,t}^k Y_{g,t}}{\sum_{g=1}^G \sum_{t\ge K+1} \eta_{g,t}^k D_{g,t-k}}.$$
Moreover, by Assumption \ref{hyp:RC_model}, for all $g$ and $t\ge K+1$,
$$Y_{g,t} = Y_{g,t}(\0_t) + \sum_{k=0}^K \beta_{g,k} D_{g,t-k}.$$
By Assumptions \ref{hyp:iid} and \ref{hyp:pt_nevertreated} and the fact that $\eta_{g,t}^k$ is a residual, we have
\begin{align*}
	E\left[\frac{\sum_{g=1}^G \sum_{t\ge K+1} \eta_{g,t}^k Y_{g,t}(\0_t)}{\sum_{g=1}^G \sum_{t\ge K+1} \eta_{g,t}^k D_{g,t-k}}\big|\bm{D}_1,...,\bm{D}_G\right] & = \frac{\sum_{g=1}^G \sum_{t\ge K+1} \eta_{g,t}^k E[Y_{g,t}(\0_t)|\bm{D}_g]}{\sum_{g=1}^G \sum_{t\ge K+1} \eta_{g,t}^k D_{g,t-k}} \\
	& = 0.
\end{align*}
Hence, by definition of $W_g^{k,k'}$,
\begin{align*}
	E\left[\widehat{\beta}_k\big|\bm{D}_1,...,\bm{D}_G\right] & = E\left[\sum_{k'=0}^K \frac{\sum_{g=1}^G \sum_{t\ge K+1} \eta_{g,t}^k D_{g,t-k'} \beta_{g,k'}}{\sum_{g=1}^G \sum_{t\ge K+1} \eta_{g,t}^k D_{g,t-k}} \big|\bm{D}_1,...,\bm{D}_G\right] \\
	& = \frac{1}{G} \sum_{g=1}^G E\left[\sum_{k'=0}^K W_g^{k,k'}\beta_{g,k} \big|\bm{D}_1,...,\bm{D}_G\right].
\end{align*}
As a result,
\begin{align*}
E\left[\widehat{\beta}_k\right] 	& = \frac{1}{G} \sum_{g=1}^G E\left[\sum_{k'=0}^K W_g^{k,k'}\beta_{g,k}\right] \\
& = E\left[\sum_{k'=0}^K W^{k,k'}\beta_k\right],
\end{align*}
where the last equality follows since groups are identically distributed. Moreover, remark that $(1/G) \sum_{g=1}^G W_g^{k,k}=1$. Since, again, groups are identically distributed, this implies that $E[W_g^{k,k}]=1$. Similarly,  since $\eta_{g,t}^k$ is a residual, $(1/G) \sum_{g=1}^G W_g^{k,k'}=0$ for all $k'\ne k$, which implies that $E[W_g^{k,k'}]=0$. Finally, if $\Cov(W^{k,k'},\beta_{k'})=0$ for all $k' \in \{0,...,K\}$,
\begin{align*}
	E\left[\widehat{\beta}_{k}\right]=& \sum_{k'=0}^K E\left[W^{k,k'} \beta_{k'}\right] \\
	= & \sum_{k'=0}^K E\left[W^{k,k'}\right] E\left[\beta_{k'}\right] \\
	= &  E\left[\beta_{k}\right].
\end{align*}


\subsection{Illustration of \eqref{eq:infinite_exp}} 
\label{sub:infinite_exp}

If $T=2$ and $K=0$, we simply have $\bm{M}=\bm{M}'=[\Delta D_2]$ and $e_0\in \text{Im}(\bm{M}')$ is equivalent to $\Delta D_2\ne 0$, in which case $f_0 = 1/\Delta D_2$. Then,
\begin{align*}
E\left[\left|e'_k\bm{M}^+(\Delta Y -\Gamma)\right| \,\big|\, e_k\in \text{Im}(\bm{M}'),\right] = & E\left[\left|\frac{\Delta Y_2 - \gamma_2}{\Delta D_2}\right| \,\big| \Delta D_2\ne 0\right] \\
= & E\left[|\beta_0 + \eps_2/\Delta D_2| \,\big| \Delta D_2\ne 0\right] 	\\
\ge & E\left[\big| |\beta_0| - |\eps_2/\Delta D_2|\big| \,| \Delta D_2\ne 0\right] \\
\ge & \big|E\left[ |\beta_0|\,\big| \Delta D_2\ne 0\right] - E\left[|\eps_2/\Delta D_2| \,\big| \Delta D_2\ne 0\right]\big|.
\end{align*}

Now, if $E[|\eps_2| \,\big| \Delta D_2]\ge s>0$ a.s. and the distribution of $|\Delta D_2|$ is absolutely continuous with respect to the Lebesgue measure, with density $f$ satisfying $\lim_{x\downarrow 0} f(x)>0$, then
\begin{align*}
E\left[|\eps_2/\Delta D_2| \,\big| \Delta D_2\ne 0\right]\ge & s\, E\left[1/|\Delta D_2| \,\big| \Delta D_2\ne 0\right] \\
= & s\, \int_0^\infty \frac{f(x)}{x}dx = \infty,
\end{align*}
where the last equality follows since $f(x)/x$ is equivalent to $[\lim_{u\downarrow 0} f(u)]/x$ as $x\to 0$.


\subsection{Proof of Theorem \ref{thm:ident_RC}} 
\label{sub:theorem_ref_thm_ident_rc}

We need to show \eqref{eq:ident_B}, and \eqref{eq:ident_B_informal} when $D_t$ is finitely supported for all $t$. Regarding the first point, first remark that by the dominated convergence theorem,
\begin{align}
\lim_{C\uparrow \infty} E\left[\beta_k\ind{\|\bm{M}'{}^+e_k\|\le C}|\, \big|\, e_k\in \text{Im}(\bm{M}')\right] =& \overline{\beta}_k, \label{eq:tcd1} \\
\lim_{C\uparrow \infty} P\left(\ind{\|\bm{M}'{}^+e_k\|\le C}|\, \big|\, e_k\in \text{Im}(\bm{M}')\right) = & 1.	\label{eq:tcd2}
\end{align}
This implies that for $C$ large enough, $P\left(e_k\in \text{Im}(\bm{M}'),\ind{\|\bm{M}'{}^+e_k\|\le C}\right)>0$. We consider such $C$ below. Now, recall that if $e_k\in \text{Im}(\bm{M}')$, then $\bm{M}'\bm{M}'{}^+ e_k=e_k$. As a result, $e'_k\bm{M}^+(\Delta Y -\Gamma) = \beta_k + e'_k\bm{M}^+\bm{\eps}$. Hence,
\begin{align*}
& E\left[\left|e'_k\bm{M}^+(\Delta Y -\Gamma)\right| \,\big|\, e_k\in \text{Im}(\bm{M}'), \|\bm{M}'{}^+e_k\|\le C\right] \\
\le & E\left[|\beta_k| \,\big|\, e_k\in \text{Im}(\bm{M}'), \|\bm{M}'{}^+e_k\|\le C\right] + M E\left[\|\bm{\eps}\| \, \big|\, e_k\in \text{Im}(\bm{M}'), \|\bm{M}'{}^+e_k\|\le C\right] < \infty.
\end{align*}
Thus, for $C$ large enough, the expectation on the right-hand side of \eqref{eq:ident_B} is well-defined. Moreover,
\begin{align*}
E\left[e'_k\bm{M}^+(\Delta Y -\Gamma) \,\big|\, e_k\in \text{Im}(\bm{M}'), \|\bm{M}'{}^+e_k\|\le C\right] & = E\left[\beta_k\,\big|\, e_k\in \text{Im}(\bm{M}'), \|\bm{M}'{}^+e_k\|\le C\right] \\
& \hspace{-0.4cm} + E\left[e'_k\bm{M}^+\bm{\eps},\big|\, e_k\in \text{Im}(\bm{M}'), \|\bm{M}'{}^+e_k\|\le C \right] \\
& = E\left[\beta_k\,\big|\, e_k\in \text{Im}(\bm{M}'), \|\bm{M}'{}^+e_k\|\le C\right] \\
& \to \overline{\beta}_k,
\end{align*}
where the last line follows by \eqref{eq:tcd1}-\eqref{eq:tcd2}.

\medskip
Turning to the second point, remark that when $D_t$ is finitely supported, $\bm{M}'$ and then $\bm{M}'{}^+e_k$ are also finitely supported. Hence, $\|\bm{M}'{}^+e_k\|$ is bounded with probability one and from what precedes,
\begin{align*}
  E\left[\left|e'_k\bm{M}^+(\Delta Y -\Gamma)\right| \,\big|\, e_k\in \text{Im}(\bm{M}')\right] = & E\left[\left|e'_k\bm{M}^+(\Delta Y -\Gamma)\right| \,\big|\, e_k\in \text{Im}(\bm{M}'), \|\bm{M}'{}^+e_k\|\le C_0\right] \\
  < & \infty,
\end{align*}
for all $C$ large enough. Then, the right-hand side of \eqref{eq:ident_B_informal} is well-defined, and the equality holds.



\bibliography{Vol1_ch11_biblio.bib}

\end{document}